\documentclass[eqsecnum,nofootinbib,11pt]{revtex4}
\usepackage{amsmath}
   \usepackage{bm}
\usepackage{amsthm,amscd}
\usepackage{mathrsfs}
\usepackage{verbatim}
\usepackage{amsfonts,amsmath,latexsym,amssymb,txfonts,bm}
\usepackage[american]{babel}
\usepackage{dsfont}
\usepackage[dvips]{graphicx}
\usepackage{subfigure}

\newcommand\bea{\begin{eqnarray}}
\newcommand\eea{\end{eqnarray}}

\begin{document}
\title{The Casimir Effect for Conical Pistons}
\author{
Guglielmo Fucci\footnote{Electronic address: Guglielmo\textunderscore Fucci@Baylor.edu} and Klaus Kirsten\footnote{Electronic address: Klaus\textunderscore Kirsten@Baylor.edu}
\thanks{Electronic address: gfucci@nmt.edu}}
\affiliation{Department of Mathematics, Baylor University, Waco, TX 76798 USA
}
\date{\today}
\vspace{2cm}
\begin{abstract}

In this paper we utilize $\zeta$-function regularization techniques in order to compute the Casimir force for massless scalar fields subject to Dirichlet and Neumann boundary conditions in the setting of the conical piston. The piston geometry is obtained by dividing the bounded generalized cone into two regions separated by
its cross section positioned at $a$ with $a\in(0,b)$ with $b>0$. We obtain expressions for the Casimir force that are valid in any dimension for both Dirichlet and Neumann boundary conditions
in terms of the spectral $\zeta$-function of the piston. As a particular case, we specify the piston to be a $d$-dimensional sphere and present explicit results for $d=2,3,4,5$.
\end{abstract}
\maketitle

\section{Introduction}

The Casimir effect is one of the most important macroscopic manifestations of the zero point energy of quantized fields under the influence of
external conditions \cite{bordag09,milton01} or in spaces with non-trivial topology. In recent years, a vast amount of literature has been produced on
the Casimir effect, which was first predicted in the seminal paper \cite{casimir48}, especially for its relevance
in nanoscale physics \cite{bordag01,bordag09,milton01}. Due to its nature, calculations of the vacuum energy lead to divergencies which
need to be regularized and subsequently renormalized. Several regularization methods exist, amongst the most important ones are frequency cutoff, point splitting and
zeta function regularization \cite{blau88,bordag01,bytsenko03,elizalde,elizalde94,milton01}. For many configurations, these techniques yield the same finite renormalized result, however the
way divergencies are removed is different in each scheme. The non-uniqueness of the removal procedure raises the question, which of them
is the physically best motivated one.
Technical and interpretational problems of this nature can actually be avoided if one considers the Casimir effect between separate objects.
In this case, the divergent part of the energy (for massless fields) depends on the heat kernel coefficient $a_{D/2}$ related to the geometry of the objects.
These coefficients, in turn, do not depend on the distance between the bodies and, hence, the Casimir force between them is free of divergencies \cite{bordag01}.
Belonging to the class of configurations for which the Casimir force has been unambiguously evaluated are pistons of certain types.

These piston configurations, introduced in \cite{cavalcanti04}, have become increasingly important because of this fact. A large variety of piston configurations and boundary conditions
have been studied throughout the literature, both at zero and at finite temperature.  Rectangular Casimir pistons with different types of boundary conditions
have been considered, for instance, in \cite{edery06,hertz05,hertz07,li97,zhai07}. Cylindrical Casimir pistons, instead, have been analyzed in \cite{barton06,marachevsky07,teo09b}.
Higher dimensional Casimir pistons with arbitrary cross sections have been studied in the setting of
Kaluza-Klein models in \cite{kirsten09,fulling09}. A generalization of these models for finite temperatures has been obtained, e.g., in \cite{teo09,teo09a}.

In the literature, Casimir piston configurations have the general geometric structure of a direct product $I\times N\times M$,
where $I$ is a closed interval of the real line, $N$ is either a compact or non-compact manifold, and $M$ is a compact manifold usually describing
the extra-dimensions. Pistons of a different geometric nature, where the local line element contains a warp factor, have been recently considered in
the framework of five and higher-dimensional Randall-Sundrum models \cite{cheng10,flachi01,rype10}. Moreover, vacuum polarization effects have been studied for
massive fermion fields in the context of the global monopole in \cite{bezerra06}.

In this paper we will focus our attention on a new type
of Casimir piston modeled on a conical manifold. The main difference of this configuration from the previous ones studied in the literature is the presence
of a singularity at the origin and the fact that the piston is a curved manifold. As a result, only for particular dimensions the force will be unambiguous.
This setup is particularly important in order to study the effects on the piston due to the presence of a geometric singularity.
The main physical interest for the study of conical manifolds lies in field theoretical models requiring an orbifold compactification \cite{fucci10,vongersdorff08}.
In fact an orbifold is defined locally by the quotient space of a smooth manifold $X$ and a discrete isometry group $G$. The action of the group on the
manifold has, in general, fixed points which are mapped to conical singularities in the quotient space. This is a topic of great interest especially in the ambit
of string theory \cite{bailin99}.

We will utilize $\zeta$-function techniques in order to obtain analytic expressions for the Casimir energy of
the conical piston. The spectrum of a self-adjoint partial differential operator on compact manifolds is discrete, bounded from below and
the eigenvalues $\lambda_{n}$ form an increasing sequence. In this case one can define the spectral $\zeta$-function associated to the operator
as
\begin{equation}\label{0}
\zeta(s)=\sum_{n=1}^\infty\lambda_{n}^{-s}\;,
\end{equation}
which is convergent for $\Re(s)>D/2$, with $D$ being the dimension of the manifold under consideration.
One can analytically continue, in a unique way, $\zeta(s)$ to a meromorphic function with only simple poles in the whole
complex plane which coincides with (\ref{0}) in its domain of convergence.

The outline of the paper is as follows. In section $2$ we describe the geometry of the conical piston and we introduce the basic objects needed for our study.
In particular, we consider two types of boundary conditions, namely Dirichlet and Neumann boundary conditions.
In the framework of $\zeta$-function regularization we obtain expressions for the Casimir force which explicitly show their dependence on the particular geometry of the piston. We specialize our general formulas to the case in which the piston is a $d$-dimensional sphere and give very explicit results in particular dimensions.
The Appendix contains a list of polynomials needed for the computation and the Conclusions point to the most important results of the article.

\section{The Conical Piston}

We will consider a particular manifold which is termed the bounded generalized cone.
The generalized cone is defined as the $D=(d+1)$-dimensional manifold $\mathscr{M}=I\times\mathscr{N}$
where $\mathscr{N}$ is the base manifold, assumed to be a smooth Riemannian manifold possibly with boundary, and $I=[0,1]\subset \mathds{R}$.
The manifold $\mathscr{M}$ is locally described by the hyperspherical metric \cite{cheeger83}
\begin{equation}\label{1}
ds^{2}=dr^{2}+r^{2}d\Sigma^{2}\;,
\end{equation}
where $d\Sigma^{2}$ represents the metric on $\mathscr{N}$ and $r\in I$.
It is known \cite{bordag96} that the curvatures on $\mathscr{M}$ and on the base $\mathscr{N}$
are conformally related as follows
\begin{eqnarray}\label{2}
R^{ij}{}_{kl}&=&\frac{1}{r^{2}}\left[\hat{R}^{ij}{}_{kl}-\left(\delta^{i}{}_{k}\delta^{j}{}_{l}-\delta^{i}{}_{l}\delta^{j}{}_{k}\right)\right]\;,\quad
R^{i}{}_{j}=\frac{1}{r^{2}}\left[\hat{R}^{i}{}_{j}-(d-1)\delta^{i}{}_{j}\right]\;,\\
R&=&\frac{1}{r^{2}}\left[\hat{R}-d(d-1)\right]\;,\nonumber
\end{eqnarray}
where $R$ and $\hat{R}$ are the curvature tensors, respectively, on $\mathscr{M}$ and $\mathscr{N}$.
It can be readily understood, from the relations (\ref{2}), that, in general, the manifold under consideration has a
singularity at the origin $r=0$.

For this type of singular Riemannian manifold the heat kernel and functional determinant of the
associated Laplace operator have been studied for massless and massive fields in \cite{bordag96,fucci10}.

Let us next describe the piston configuration that arises from the generalized cone.
Consider a cross section of the manifold $\mathscr{M}$ positioned at the point $r=a$ with
$a\in(0,b)$ and $b>0$. We will denote this cross section by $\mathscr{N}_a$. Note that for any $x$ and $y$ in the interval $(0,b)$, $\mathscr{N}_x$ and $\mathscr{N}_y$ are diffeomorphic. The $d$-dimensional manifold $\mathscr{N}_a$
naturally divides the manifold $\mathscr{M}$ into two distinct regions: region $I$ represented by $M_{I}=[0,a]\times \mathscr{N}$ and region $II$ represented by
$M_{II}=(a,b]\times \mathscr{N}$ where the first, $M_{I}$, contains the singularity at $r=0$. $M_{I}$ and $M_{II}$
represent two $D$-dimensional manifolds with boundary, where $\partial M_{I}=\{0\}\cup\mathscr{N}_{a}$ and $\partial M_{II}=\mathscr{N}_{a}\cup\mathscr{N}_{b}$.

Clearly, the generalized cone
$\mathscr{M}$ is obtained from the union of $M_{I}$ and $M_{II}$ along their common boundary $\mathscr{N}_a$,
more precisely
\begin{displaymath}
\mathscr{M}=M_{I}\,\cup_{\mathscr{N}_a}M_{II}\;.
\end{displaymath}
The configuration that we have described above is a conical piston, where the piston itself
is modeled by the cross section $\mathscr{N}_a$ of $\mathscr{M}$ at the point $r=a$.

We will consider, in what follows, the Laplace operator $\Delta_\mathscr{M}$ on $\mathscr{M}$ acting
on the space $\mathscr{L}^{2}(\mathscr{M})$ of square integrable scalar functions on the generalized cone.
The starting point of our analysis is the following eigenvalue problem
\begin{equation}\label{3}
\left(-\Delta_{\mathscr{M}}+m^{2}\right)\varphi=\alpha^{2}\varphi\;,
\end{equation}
where we have introduced the spectral parameter (mass) $m$ in order to make some of the subsequent integrals well defined.
At the end of the calculation the limit $m\to 0$ will be taken, giving results for massless scalar fields.
In hyperspherical coordinates the Laplacian $\Delta_{\mathscr{M}}$ takes the form of
a Bessel type operator as follows
\begin{equation}\label{4}
\Delta_{\mathscr{M}}=\frac{\partial^{2}}{\partial r^{2}}+\frac{d}{r}\frac{\partial}{\partial r}+\frac{1}{r^{2}}\Delta_{\mathscr{N}}\;,
\end{equation}
with $\Delta_\mathscr{N}$ denoting the Laplace operator on the manifold $\mathscr{N}$.

The idea is to solve the eigenvalue problem (\ref{3}) in both region $I$ and region $II$. The fields
in one region are independent of the fields in the other region and therefore the corresponding spectral problems are independent.

Region $I$, corresponding to the manifold $M_{I}$, contains the conical singularity at $r=0$.
We require the solution of (\ref{3}) to be regular at the origin, and we obtain
\begin{equation}\label{5}
\varphi_{I}=r^{\frac{1-d}{2}}J_{\nu}(\gamma_{I}r )\Phi(\Omega)\;,
\end{equation}
where $J_{\nu}$ is the Bessel function of the first kind and we have set $\alpha_{I}^{2}=\gamma_{I}^{2}+m^{2}$. The angular functions $\Phi(\Omega)$ represent
the hyperspherical harmonics on $\mathscr{N}$ satisfying the eigenvalue equation
\begin{equation}\label{6}
\Delta_{\mathscr{N}}\Phi(\Omega)=-\lambda^{2}\Phi(\Omega)\;.
\end{equation}

Region $II$, represented by the manifold $M_{II}$, does not contain the conical singularity at $r=0$. Therefore,
a general solution to the eigenvalue problem (\ref{3}) will be a linear combination of Bessel functions
of the first and second kind as follows
\begin{equation}\label{7}
\varphi_{II}=r^{\frac{1-d}{2}}\Big[A\,J_{\nu}(\gamma_{II}r)+B\,Y_{\nu}(\gamma_{II}r)\Big]\Phi(\Omega)\;,
\end{equation}
where $A$ and $B$ are arbitrary constants, and we have set $\alpha_{II}^{2}=\gamma_{II}^{2}+m^{2}$.
The index $\nu$ of the Bessel function is easily found by substituting the general solutions (\ref{5}) and (\ref{7}) into (\ref{3}).
By taking into account the angular relation (\ref{6}), one finds that the radial differential equation, in both regions, is satisfied
if the following holds
\begin{equation}\label{8}
\nu^{2}=\lambda^{2}+\frac{(1-d)^{2}}{4}\;.
\end{equation}

We would like to introduce, at this point, the $\zeta$-function associated to the spectrum of the operator (\ref{4}) in regions $I$ and $II$. Since the form of the
respective $\zeta$-functions is the same, we will utilize a unified notation and define
\begin{equation}\label{9}
\zeta_{i}(s)=\sum_{\gamma_{i}}(\gamma_{i}^{2}+m^{2})^{-s}\;,
\end{equation}
where $i$ represents either $I$ or $II$ and we will assume that no negative eigenvalues occur so that we can use the standard branch cut of the logarithm.
It is clear that the $\zeta$-function associated with the piston on the manifold $\mathscr{M}$ is obtained
by adding the contributions from both regions,
\begin{equation}\label{9a}
\zeta_{\mathscr{M}}(s)=\zeta_{I}(s)+\zeta_{II}(s)\;.
\end{equation}

In the following, we will express the $\zeta$-function on $M_{I}$ and $M_{II}$ in terms of the zeta function $\zeta_{\mathscr{N}}$ of the manifold $\mathscr{N}$ \cite{cheeger83}
defined as
\begin{equation}\label{10}
\zeta_{\mathscr{N}}(s)=\sum_{\nu} d(\nu)\nu^{-2s}\;,
\end{equation}
where $d(\nu)$ is the degeneracy of the scalar harmonics $\Phi(\Omega)$ on $\mathscr{N}$. This definition will serve the purpose
of keeping the manifold $\mathscr{N}$ unspecified throughout the calculations without jeopardizing the possibility to impose boundary conditions \cite{bordag96}.

It is useful to mention that in the framework of $\zeta$-function regularization the Casimir energy is defined as follows \cite{bordag01,bordag09,elizalde,elizalde94,kirsten01},
\begin{equation}\label{0a}
E_{\textrm{Cas}}=\lim_{\alpha\to 0}\frac{\mu^{2\alpha}}{2}\zeta_{\mathscr{M}}\left(\alpha-\frac{1}{2}\right)\;,
\end{equation}
where $\mu$ represents an arbitrary parameter with the dimension of a mass. By expanding the previous expression about $\alpha=0$, one obtains
a result which manifestly shows the structure of $E_{\textrm{Cas}}$, namely
\begin{equation}\label{41a}
E_{\textrm{Cas}}=\frac{1}{2}\textrm{FP}\zeta_{\mathscr{M}}\left(-\frac{1}{2}\right)+\frac{1}{2}\left(\frac{1}{\alpha}+\ln\mu^{2}\right)\textrm{Res}\,\zeta_{\mathscr{M}}\left(-\frac{1}{2}\right)+O(\alpha)\;,
\end{equation}
where $\textrm{Res}$ denotes the residue of the function and $\textrm{FP}$ its finite part.
The last formula indicates that the Casimir energy has an ambiguity proportional to the heat kernel coefficient $a_{D/2}$ of the operator $-\Delta_{\mathscr{M}} + m^2$, which, in turn, is
given by the residue of $\zeta_{\mathscr{M}}$ at $s=-1/2$.

In the following calculations it will therefore be sufficient to compute the residue and finite part of the spectral $\zeta$-function
on the manifold $\mathscr{M}$ at the point $s=-1/2$ to obtain $E_{\textrm{Cas}}$.
When studying traditional pistons, only the finite part of the spectral $\zeta$-function depends on the position of the piston $a$. The force
is then easily obtained from $E_{\textrm{Cas}}$ by exploiting the formula
\begin{equation}\label{10a}
F_{\textrm{Cas}}(a)=-\frac{\partial}{\partial a}E_{\textrm{Cas}}(a)\;,
\end{equation}
which gives an unambiguous result. The situation for the conical piston is, however, different. In this case
the residue of the spectral $\zeta$-function at $s=-1/2$ will be, in general, dependent on the position $a$ of the piston, as it will be shown later.
The resulting force will be obtained from (\ref{41a}) in the form
\begin{equation}\label{10b}
F_{\textrm{Cas}}(a)=-\frac{1}{2}\frac{\partial}{\partial a}\textrm{FP}\zeta_{\mathscr{M}}\left(-\frac{1}{2},a\right)-\frac{1}{2}\left(\frac{1}{\alpha}+\ln\mu^{2}\right)\frac{\partial}{\partial a}\textrm{Res}\,\zeta_{\mathscr{M}}\left(-\frac{1}{2},a\right)+O(\alpha)\;.
\end{equation}
It is thus clear that an unambiguous prediction of the force cannot be obtained in general in this setting.
However, as we will see, if the piston $\mathscr{N}$ is an even-dimensional manifold without boundary, the resulting force will
be free of divergencies.

In order to explicitly compute the spectral $\zeta$-functions in the two regions we need to impose specific boundary conditions,
which, in turn, will provide implicit equations for the eigenvalues. In this work we will consider two types of boundary conditions, namely
Dirichlet and Neumann boundary conditions.

\section{Dirichlet Boundary Conditions}

In this section we will study the conical piston endowed with Dirichlet boundary conditions. In region $I$ we impose Dirichlet boundary conditions on the piston $\mathscr{N}_{a}$ positioned at $r=a$,
which gives the following implicit equation for the eigenvalues $\gamma_{I}$
\begin{equation}\label{11}
J_{\nu}(\gamma_{I}a)=0\;.
\end{equation}
In region $II$ we impose Dirichlet boundary conditions on $\partial M_{II}$ to obtain the linear system of equations
\begin{equation}\label{12}
\left\{ \begin{array}{l}
A\,J_{\nu}(\gamma_{II}a)+B\,Y_{\nu}(\gamma_{II}a)=0\\
A\,J_{\nu}(\gamma_{II}b)+B\,Y_{\nu}(\gamma_{II}b)=0\;.\\
\end{array} \right.
\end{equation}
The above linear system has non-trivial solutions if the following equation is satisfied
\begin{equation}\label{13}
J_{\nu}(\gamma_{II}a)Y_{\nu}(\gamma_{II}b)-J_{\nu}(\gamma_{II}b)Y_{\nu}(\gamma_{II}a)=0\;.
\end{equation}
Since the index $\nu$ is real and $a$ is positive, equation (\ref{13}) has only real and simple solutions, which implicitly determines
the eigenvalues $\gamma_{II}$ in region $II$.

The starting point of our analysis is the representation of the spectral zeta function $\zeta_{i}$ in terms of a
contour integral in the complex plane valid for $\Re(s)>(d+1)/2$ \cite{bordag96,bordag96a,bordag96b,esposito97,kirsten01}. By following the standard procedure,
in region $I$ we obtain
\begin{equation}\label{14}
\zeta_{I}(s,a)=\sum_\nu d(\nu)\frac{1}{2\pi i}\int_{\Gamma}dk\left[k^{2}+m^{2}\right]^{-s}\frac{\partial}{\partial k}\ln \left[k^{-\nu} J_{\nu}(ka)\right]\;,
\end{equation}
where $m$, as mentioned, is a mass which will be sent to zero at the end of the calculation, and $\Gamma$ is a contour
that encircles all the zeroes of $J_{\nu}(ka)$ on the positive real axis in the counterclockwise direction. The term $k^{-\nu}$
has been introduced in order to avoid contributions coming from the origin in the subsequent contour deformation.
Since the origin is not included in the contour the additional term will not change the result of the integral \cite{kirsten01}.
In region $II$ we have a similar representation, namely
\begin{equation}\label{15}
\zeta_{II}(s,a)=\sum_\nu d(\nu)\frac{1}{2\pi i}\int_{\Gamma'}d\kappa\left[\kappa^{2}+m^{2}\right]^{-s}\frac{\partial}{\partial \kappa}
\ln\left[J_{\nu}(\kappa a)Y_{\nu}(\kappa b)-J_{\nu}(\kappa b)Y_{\nu}(\kappa a)\right]\;,
\end{equation}
here $\Gamma'$ is a contour
that encircles all the zeroes of $J_{\nu}(\kappa a)Y_{\nu}(\kappa b)-J_{\nu}(\kappa b)Y_{\nu}(\kappa a)$ on the positive real axis in the counterclockwise direction.

By deforming the contour of integration in (\ref{14}) to the imaginary axis and by performing the change of variables $k\to \nu k/a$ we obtain for region $I$
\begin{equation}
\zeta_{I}(s,a)=\sum_\nu d(\nu)\zeta^{\nu}_{I}(s,a)\;,
\end{equation}
where
\begin{equation}\label{16}
\zeta^{\nu}_{I}(s,a)=\frac{\sin(\pi s)}{\pi}\int_{\frac{ma}{\nu}}^{\infty}dk\left[\frac{\nu^{2}k^{2}}{a^2}-m^{2}\right]^{-s}\frac{\partial}{\partial k}\ln \left[k^{-\nu}I_{\nu}(\nu k)\right]\;.
\end{equation}
An analogous deformation to the imaginary axis of the contour $\Gamma'$ in (\ref{15}) and the change of variable $\kappa\to \kappa\nu$ leads to the expression
\begin{equation}
\zeta_{II}(s,a,b)=\sum_\nu d(\nu)\zeta^{\nu}_{II}(s,a,b)\;,
\end{equation}
with
\begin{equation}\label{17a}
\zeta^{\nu}_{II}(s,a,b)=\frac{\sin(\pi s)}{\pi}\int_{\frac{m}{\nu}}^{\infty}d\kappa\left[(\nu\kappa)^{2}-m^{2}\right]^{-s}\frac{\partial}{\partial \kappa}\ln \left[K_{\nu}(\nu\kappa a)I_{\nu}(\nu\kappa b)-K_{\nu}(\nu\kappa b)I_{\nu}(\nu\kappa a)\right]\;.
\end{equation}
The results (\ref{16}) and (\ref{17a}) are well defined in the strip $1/2<\Re(s)<1$, and have been obtained by exploiting the formulas for Bessel functions
of imaginary argument \cite{gradshtein07}.

Let us for the moment focus our attention to the spectral $\zeta$-function in region $I$. In order to analytically continue
the result to values of $s$ for which $\Re(s)<1/2$, we follow the methods of \cite{bordag96,bordag96a,bordag96b,kirsten01} and
utilize the asymptotic expansion of the modified Bessel functions $I_{\nu}(k)$ for
$\nu\to\infty$ and for $z=k/\nu$ fixed. In detail \cite{olver54,erdelyi53}
\begin{equation}\label{17}
I_{\nu}(\nu z)\sim\frac{1}{\sqrt{2\pi\nu}}\frac{e^{\nu\eta}}{(1+z^{2})^{1/4}}\left[1+\sum_{k=1}^{\infty}\frac{u_{k}(t)}{\nu^{k}}\right]\;,
\end{equation}
where the polynomials $u_{k}(t)$ are determined by the recurrence relation
\begin{equation}\label{18}
u_{k+1}(t)=\frac{1}{2}t^{2}(1-t^{2})u_{k}^{\prime}(t)+\frac{1}{8}\int_{0}^{t}d\tau(1-5\tau^{2})u_{k}(\tau)\;,
\end{equation}
with $u_{0}(t)=1$ and
\begin{equation}\label{19}
t=\frac{1}{\sqrt{1+z^{2}}}\;,\qquad \eta=\sqrt{1+z^{2}}+\ln\left[\frac{z}{1+\sqrt{1+z^{2}}}\right]\;.
\end{equation}
The desired analytic continuation of (\ref{16}) is obtained by adding and subtracting $N$ leading terms of the uniform asymptotic expansion (\ref{17}). One finds \cite{bordag96,bordag96a,bordag96b,fucci10,kirsten01}
\begin{equation}\label{20}
\zeta_{I}(s,a)=Z_{I}(s,a)+\sum_{i=-1}^{N}A^{(I)}_{i}(s,a)\;,
\end{equation}
where, once the limit $m\to 0$ is performed, we have
\begin{eqnarray}\label{21}
A^{(I)}_{-1}(s,a)&=&\frac{a^{2s}}{4\sqrt{\pi}}\,\frac{\Gamma\left(s-\frac{1}{2}\right)}{\Gamma(s+1)}\zeta_{\mathscr{N}}\left(s-\frac{1}{2}\right)\;,\\
A^{(I)}_{0}(s,a)&=&-\frac{a^{2s}}{4}\zeta_{\mathscr{N}}\left(s\right)\;,\\
A^{(I)}_{i}(s,a)&=&-\frac{a^{2s}}{\Gamma(s)}\zeta_{\mathscr{N}}\left(s+\frac{i}{2}\right)\sum_{b=0}^{i}x_{i,b}\frac{\Gamma\left(s+b+\frac{i}{2}\right)}{\Gamma\left(b+\frac{i}{2}\right)}\;.\label{21a}
\end{eqnarray}
The remaining term, namely $Z_{I}(s,a)$, represents, by construction, an analytic function for $\Re(s)>(d-1-N)/2$ which is defined as
\begin{equation}
Z_I(s,a)=\sum_{\nu} d(\nu)Z^{(I)}_{\nu}(s,a)\;,
\end{equation}
with
\begin{eqnarray}\label{22}
Z^{(I)}_{\nu}(s,a)&=&a^{2s}\nu^{-2s}\frac{\sin(\pi s)}{\pi}\int_{0}^{\infty}dk\,k^{-2s}\frac{\partial}{\partial k}
\Bigg\{\ln\left[k^{-\nu}I_{\nu}(k\nu)\right]
-\ln\left[\frac{k^{-\nu}}{\sqrt{2\pi\nu}}\frac{e^{\nu\eta}}{(1+k^{2})^{1/4}}\right]
-\sum_{n=1}^{N}\frac{D_{n}(t)}{\nu^{n}}\Bigg\}\;.\;\;\;\;\;\;
\end{eqnarray}
The terms $D_{n}(t)$ appearing in (\ref{20}) are defined through the cumulant expansion \cite{bordag96,bordag96a,bordag96b,fucci10,kirsten01}
\begin{equation}\label{23}
\ln\left[1+\sum_{k=1}^{\infty}\frac{u_{k}(t)}{\nu^{k}}\right]\sim\sum_{n=1}^{\infty}\frac{D_{n}(t)}{\nu^{n}}\;,
\end{equation}
and have the polynomial structure
\begin{equation}\label{24}
D_{n}(t)=\sum_{i=0}^{n}x_{i,n}t^{n+2i}\;.
\end{equation}
For a list of the first polynomials $D_{n}(t)$ the reader is referred to the appendix.

The spectral $\zeta$-function in region $II$, given by the integral (\ref{17a}), can be conveniently rewritten
as a sum of three distinct terms
\begin{eqnarray}\label{25}
\zeta^{\nu}_{II}(s,a,b)&=&\frac{\sin(\pi s)}{\pi}\int_{\frac{ma}{\nu}}^{\infty}d\kappa\left[\frac{\nu^2\kappa^2}{a^2}-m^{2}\right]^{-s}\frac{\partial}{\partial \kappa}\ln \left[\kappa^{\nu}K_{\nu}(\nu\kappa)\right]\nonumber\\
&+&\frac{\sin(\pi s)}{\pi}\int_{\frac{m b}{\nu}}^{\infty}d\kappa\left[\frac{\nu^{2}\kappa^{2}}{b^{2}}-m^{2}\right]^{-s}\frac{\partial}{\partial \kappa}\ln \left[\kappa^{-\nu}I_{\nu}(\nu\kappa)\right]+\mathscr{F}^{\nu}_{\mathcal{D}}(s,a,b)\;,
\end{eqnarray}
where $\mathscr{F}^{\nu}_{\mathcal{D}}(s,a,b)$ has the following integral representation, once the limit $m\to 0$ has been taken,
\begin{equation}\label{26}
\mathscr{F}^{\nu}_{\mathcal{D}}(s,a,b)=\nu^{-2s}\frac{\sin(\pi s)}{\pi}\int_{0}^{\infty}d\kappa\,\kappa^{-2s}\frac{\partial}{\partial \kappa}\ln
\left[1-\frac{K_{\nu}(\nu\kappa b)I_{\nu}(\nu\kappa a)}{K_{\nu}(\nu\kappa a)I_{\nu}(\nu\kappa b)}\right]\;.
\end{equation}
The first two integrals in (\ref{25}) need to be analytically continued to values of $s$ for which $\Re(s)<1/2$.
In complete analogy to the methods used for the analytic continuation of $\zeta_{I}(s,a)$, we consider the uniform asymptotic expansion
of the modified Bessel functions $K_{\nu}(k)$ for
$\nu\to\infty$ and for $z=k/\nu$ fixed, namely \cite{olver54,erdelyi53}
\begin{equation}\label{27}
K_{\nu}(\nu z)\sim\sqrt{\frac{\pi}{2\nu}}\frac{e^{-\nu\eta}}{(1+z^{2})^{1/4}}\left[1+\sum_{k=1}^{\infty}(-1)^{k}\frac{u_{k}(t)}{\nu^{k}}\right]\;.
\end{equation}
By adding and subtracting $N$ leading terms of the uniform asymptotic expansion (\ref{27}) and (\ref{17}), we obtain, for $\zeta_{II}(s,a,b)$, the following
expression
\begin{equation}\label{28}
\zeta_{II}(s,a,b)=Z_{II}(s,a,b)+\mathscr{F}_{\mathcal{D}}(s,a,b)+\sum_{i=-1}^{N}A^{(II)}_{i}(s,a,b)\;,
\end{equation}
where
\begin{equation}
  \mathscr{F}_{\mathcal{D}}(s,a,b)=\sum_{\nu}d(\nu)\mathscr{F}^{\nu}_{\mathcal{D}}(s,a,b)\;,
\end{equation}
\begin{equation}\label{29}
A^{(II)}_{i}(s,a)=(-1)^{i}A^{(I)}_{i}(s,a)+A_{i}^{(I)}(s,b)\;,
\end{equation}
and $Z_{II}(s,a,b)$ is an analytic function for $\Re(s)>(d-1-N)/2$ given by the expression
\begin{eqnarray}\label{30}
Z_{II}(s,a,b)=\sum_{\nu}d(\nu)Z_{\nu}^{(II)}(s,a,b)\;,
\end{eqnarray}
with
\begin{eqnarray}\label{31}
Z_{\nu}^{(II)}(s,a,b)&=&a^{2s}\nu^{-2s}\frac{\sin(\pi s)}{\pi}\int_{0}^{\infty}d\kappa\,\kappa^{-2s}\frac{\partial}{\partial \kappa}
\Bigg\{\ln\left[\kappa^{\nu}K_{\nu}(\kappa\nu)\right]
-\ln\left[\sqrt{\frac{\pi}{2\nu}}\frac{\kappa^{\nu}e^{-\nu\eta}}{(1+\kappa^{2})^{1/4}}\right]
-\sum_{n=1}^{N}(-1)^{n}\frac{D_{n}(t)}{\nu^{n}}\Bigg\}\nonumber\\
&+&b^{2s}\nu^{-2s}\frac{\sin(\pi s)}{\pi}\int_{0}^{\infty}d\kappa\,\kappa^{-2s}\frac{\partial}{\partial \kappa}
\Bigg\{\ln\left[\kappa^{-\nu}I_{\nu}(\kappa\nu)\right]
-\ln\left[\frac{\kappa^{-\nu}e^{-\nu\eta}}{\sqrt{2\pi\nu}(1+\kappa^{2})^{1/4}}\right]
-\sum_{n=1}^{N}\frac{D_{n}(t)}{\nu^{n}}\Bigg\}\;.\;\;\;\;\;\;\;\;\;\;\;
\end{eqnarray}

The function defined in (\ref{26}) can actually be studied in more detail \cite{flachi01,teo10}. As $\kappa$ approaches $0$ the integral in $\mathscr{F}^{\nu}_{\mathcal{D}}(s,a,b)$
represents a well defined function in the region $\Re(s)<1/2$. As $\kappa\to\infty$, we utilize the uniform asymptotic expansion of the modified
Bessel functions (\ref{17}) and (\ref{27}) to obtain
\begin{equation}\label{32}
\frac{K_{\nu}(\nu\kappa b)I_{\nu}(\nu\kappa a)}{K_{\nu}(\nu\kappa a)I_{\nu}(\nu\kappa b)}\sim
\exp\left\{-2\nu\left[\eta (b\kappa) - \eta (a \kappa) \right]\right\}\;.
\end{equation}
Since $a<b$, we have $\eta (a\kappa) < \eta (b\kappa )$ and the contributions of the
expression (\ref{32}) as $\kappa\to\infty$ are exponentially suppressed. These remarks allow us to conclude that $\mathscr{F}^{\nu}_{\mathcal{D}}(s,a,b)$
defines an analytic function for $\Re(s)<1/2$ and we can safely set $s=-1/2$ in $\mathscr{F}_{\mathcal{D}}(s,a,b)$.

In order to compute the Casimir energy for the configuration under consideration, we need to evaluate $\zeta_{I}(s,a,b)$
and $\zeta_{II}(s,a,b)$ in the neighborhood of $s=-1/2$. We would like to point out that for the explicit calculations that
will follow it will be sufficient to subtract the first $D$ terms of the asymptotic expansions (\ref{17}) and (\ref{27}) \cite{kirsten01}.
Therefore for the rest of the paper we will set $N=D$. In this way the functions $Z_{I}(s,a,b)$ and $Z_{II}(s,a,b)$ in (\ref{22}) and (\ref{31}) are, by construction,
analytic in the strip $-1<\Re(s)<1/2$ and, consequently, will not contribute to the residue of the pole of the $\zeta$-functions at $s=-1/2$.
The evaluation of the functions $A^{(I)}_{i}(s,a)$ at the point $s=-1/2$ needs more care since the pole structure of $\zeta_{I}$ and $\zeta_{II}$ is encoded in these
terms. To obtain a systematic expansion, we set $s=-1/2+\alpha$ and expand the resulting expressions for $A^{(I)}_{i}(s,a)$ about $\alpha=0$.
From the general theory of spectral $\zeta$-functions \cite{gilkey95,kirsten01} we have the following expansion in terms of the variable $\alpha$
\begin{equation}\label{33}
\zeta_{\mathscr{N}}(\alpha-1)=\zeta_{\mathscr{N}}(-1)+\alpha\zeta_{\mathscr{N}}'(- 1)+O(\alpha^{2})\;,\quad \zeta_{\mathscr{N}}(\alpha)=\zeta_{\mathscr{N}}(0)+\alpha\zeta_{\mathscr{N}}'(0)+O(\alpha^{2})\;,
\end{equation}
\begin{equation}\label{34}
\zeta_{\mathscr{N}}\left(\alpha-\frac{1}{2}\right)=\frac{1}{\alpha}\textrm{Res}\,\zeta_{\mathscr{N}}\left(-\frac{1}{2}\right)+\textrm{FP}\,\zeta_{\mathscr{N}}\left(-\frac{1}{2}\right)+O(\alpha)\;,
\end{equation}
and, for all $d+1\geq i\geq 2$,
\begin{equation}\label{35}
\zeta_{\mathscr{N}}\left(\alpha+\frac{i-1}{2}\right)=\frac{1}{\alpha}\textrm{Res}\,\zeta_{\mathscr{N}}\left(\frac{i-1}{2}\right)+\textrm{FP}\,\zeta_{\mathscr{N}}\left(\frac{i-1}{2}\right)+O(\alpha)\;.
\end{equation}

By utilizing the formulas above we are able to obtain
\begin{equation}\label{36}
A^{(I)}_{-1}\left(\alpha-\frac{1}{2},a\right)=-\frac{1}{\alpha}\frac{\zeta_{\mathscr{N}}(-1)}{4\pi a}-\frac{1}{4\pi a}\left[\Big(\ln a^{2}+2\ln 2+1\Big)\zeta_{\mathscr{N}}(-1)+\zeta'_{\mathscr{N}}(-1)\right]+O(\alpha)\;,
\end{equation}
\begin{equation}\label{37}
A^{(I)}_{0}\left(\alpha-\frac{1}{2},a\right)=-\frac{1}{4a\alpha}\textrm{Res}\,\zeta_{\mathscr{N}}\left(-\frac{1}{2}\right)-\frac{1}{4a}\left[\textrm{FP}\,\zeta_{\mathscr{N}}\left(-\frac{1}{2}\right)+\ln a^{2}\,\textrm{Res}\,\zeta_{\mathscr{N}}\left(-\frac{1}{2}\right)\right]+O(\alpha)\;.
\end{equation}
The term in (\ref{21a}) with $i=1$ should be treated separately due to the fact that $s=0$ is a regular point of the spectral $\zeta$-function $\zeta_{\mathscr{N}}(s)$. More explicitly one has
\begin{equation}\label{38}
A^{(I)}_{1}\left(\alpha-\frac{1}{2},a\right)=\frac{1}{\alpha}\left[\frac{1}{16\pi a}\zeta_{\mathscr{N}}(0)\right]+\frac{1}{16\pi a}\left[\zeta'_{\mathscr{N}}(0)+\zeta_{\mathscr{N}}(0)\left(\ln a^{2}+2\ln 2-\frac{16}{3}\right)\right]+O(\alpha)\;.
\end{equation}
For $d+1\geq i\geq 2$ one arrives, instead, at the following expression
\begin{eqnarray}\label{39}
A^{(I)}_{i}\left(\alpha-\frac{1}{2},a\right)&=&\frac{1}{\alpha}\left[\frac{\omega_{i}}{2a\sqrt{\pi}}\textrm{Res}\,\zeta_{\mathscr{N}}\left(\frac{i-1}{2}\right)\right]+\frac{1}{2a\sqrt{\pi}}\Bigg[\omega_{i}\textrm{FP}\,\zeta_{\mathscr{N}}\left(\frac{i-1}{2}\right)\nonumber\\
&+&\omega_{i}\left(\ln a^{2}+\gamma+2\ln 2-2\right)\textrm{Res}\,\zeta_{\mathscr{N}}\left(\frac{i-1}{2}\right)+\Omega_{i}\textrm{Res}\,\zeta_{\mathscr{N}}\left(\frac{i-1}{2}\right)\Bigg]+O(\alpha)\;,
\end{eqnarray}
where $\gamma$ is the Euler-Mascheroni constant and we have defined the quantities
\begin{equation}\label{40}
\omega_{i}=\sum_{p=0}^{i}x_{i,p}\frac{\Gamma\left(p+\frac{i-1}{2}\right)}{\Gamma\left(p+\frac{i}{2}\right)}\;,\qquad
\Omega_{i}=\sum_{p=0}^{i}x_{i,p}\frac{\Gamma\left(p+\frac{i-1}{2}\right)}{\Gamma\left(p+\frac{i}{2}\right)}\Psi\left(p+\frac{i-1}{2}\right)\;,
\end{equation}
with $\Psi(x)$ being the logarithmic derivative of the Gamma function. From the definition (\ref{9a}), we have the following expression for $\zeta_{\mathscr{M}}(s)$ for
Dirichlet boundary conditions
\begin{eqnarray}\label{41}
\zeta_{\mathscr{M}}\left(\alpha-\frac{1}{2},a,b\right)&=&Z_{I}\left(-\frac{1}{2},a\right)+Z_{II}\left(-\frac{1}{2},a,b\right)+\mathscr{F}_{\mathcal{D}}\left(-\frac{1}{2},a,b\right)\nonumber\\
&+&2\sum_{i=0}^{[D/2]}A^{(I)}_{2i}\left(-\frac{1}{2}+\alpha,a\right)+\sum_{i=-1}^{D}A^{(I)}_{i}\left(-\frac{1}{2}+\alpha,b\right)\;,
\end{eqnarray}
where $[x]$ represents the integer part of $x$.

According to (\ref{41a}), we need to extract the residue and finite part of the $\zeta$-function above in order to evaluate the Casimir energy.
From equations (\ref{36})-(\ref{39}) and, by recalling (\ref{29}), we explicitly have
\begin{eqnarray}\label{42}
\textrm{Res}\,\zeta_{\mathscr{M}}\left(-\frac{1}{2},a,b\right)&=&-\frac{1}{2}\left(\frac{1}{a}+\frac{1}{2b}\right)\textrm{Res}\,\zeta_{\mathscr{N}}\left(-\frac{1}{2}\right)+\frac{1}{a\sqrt{\pi}}\sum_{i=1}^{[D/2]}\omega_{2i}
\textrm{Res}\,\zeta_{\mathscr{N}}\left(\frac{2i-1}{2}\right)-\frac{1}{4\pi b}\zeta_{\mathscr{N}}\left(-1\right)\nonumber\\
&+&\frac{1}{16\pi b}\zeta_{\mathscr{N}}\left(0\right)+\frac{1}{2\sqrt{\pi} b}\sum_{i=2}^{D}\omega_{i}\textrm{Res}\,\zeta_{\mathscr{N}}\left(\frac{i-1}{2}\right)\;,
\end{eqnarray}
while for the finite part we get
\begin{eqnarray}\label{43}
\textrm{FP}\,\zeta_{\mathscr{M}}\left(-\frac{1}{2},a,b\right)&=&Z_{I}\left(-\frac{1}{2},a\right)+Z_{II}\left(-\frac{1}{2},a,b\right)+\mathscr{F}_{\mathcal{D}}\left(-\frac{1}{2},a,b\right)-\frac{1}{2a}\left[\textrm{FP}\,\zeta_{\mathscr{N}}\left(-\frac{1}{2}\right)+\ln a^{2}\,\textrm{Res}\,\zeta_{\mathscr{N}}\left(-\frac{1}{2}\right)\right]\nonumber\\
&+&\frac{1}{a\sqrt{\pi}}\sum_{i=1}^{[D/2]}\Bigg[\omega_{2i}\textrm{FP}\,\zeta_{\mathscr{N}}\left(\frac{2i-1}{2}\right)
+\omega_{2i}\left(\ln a^{2}+\gamma+2\ln 2-2\right)\textrm{Res}\,\zeta_{\mathscr{N}}\left(\frac{2i-1}{2}\right)\nonumber\\
&+&\Omega_{2i}\textrm{Res}\,\zeta_{\mathscr{N}}\left(\frac{2i-1}{2}\right)\Bigg]-\frac{1}{4\pi b}\left[\Big(2\ln 2+1\Big)\zeta_{\mathscr{N}}(-1)+\zeta'_{\mathscr{N}}(-1)\right]-\frac{1}{4b}\textrm{FP}\,\zeta_{\mathscr{N}}\left(-\frac{1}{2}\right)\nonumber\\
&+&\frac{1}{16\pi b}\left[\zeta'_{\mathscr{N}}(0)+\zeta_{\mathscr{N}}(0)\left(2\ln 2-\frac{16}{3}\right)\right]+\frac{1}{2\sqrt{\pi} b}\sum_{i=2}^{D}\Bigg[\omega_{i}\textrm{FP}\,\zeta_{\mathscr{N}}\left(\frac{i-1}{2}\right)\nonumber\\
&+&\omega_{i}\left(\gamma+2\ln 2-2\right)\textrm{Res}\,\zeta_{\mathscr{N}}\left(\frac{i-1}{2}\right)+\Omega_{i}\textrm{Res}\,\zeta_{\mathscr{N}}\left(\frac{i-1}{2}\right)\Bigg]\;.
\end{eqnarray} From formula (\ref{10b}) we can then write an explicit expression for the force on the piston
when Dirichlet boundary conditions are imposed
\begin{eqnarray}\label{44}
F^{\textrm{Dir}}_{\textrm{Cas}}(a,b)&=&-\frac{1}{2}Z'_{I}\left(-\frac{1}{2},a\right)-\frac{1}{2}Z'_{II}\left(-\frac{1}{2},a\right)-\frac{1}{2}\mathscr{F}'_{\mathcal{D}}\left(-\frac{1}{2},a,b\right)+\frac{1}{4a^{2}}\left[(2-\ln a^{2})\textrm{Res}\,\zeta_{\mathscr{N}}\left(-\frac{1}{2}\right)-\textrm{FP}\,\zeta_{\mathscr{N}}\left(-\frac{1}{2}\right)\right]\nonumber\\
&+&\frac{1}{2\sqrt{\pi}a^{2}}\sum_{i=1}^{[D/2]}\Bigg[\omega_{2i}\textrm{FP}\,\zeta_{\mathscr{N}}\left(\frac{2i-1}{2}\right)
-\omega_{2i}\left(4-\ln a^{2}-\gamma-2\ln 2\right)\textrm{Res}\,\zeta_{\mathscr{N}}\left(\frac{2i-1}{2}\right)\nonumber\\
&+&\Omega_{2i}\textrm{Res}\,\zeta_{\mathscr{N}}\left(\frac{2i-1}{2}\right)\Bigg]-\frac{1}{4a^{2}}\left(\frac{1}{\alpha}+\ln\mu^{2}\right)\Bigg[
\textrm{Res}\,\zeta_{\mathscr{N}}\left(-\frac{1}{2}\right)-\frac{2}{\sqrt{\pi}}\sum_{i=1}^{[D/2]}\omega_{2i}\textrm{Res}\,\zeta_{\mathscr{N}}\left(\frac{2i-1}{2}\right)\Bigg]+O(\alpha)\;,\nonumber\\
\end{eqnarray}
where the prime indicates differentiation with respect to the variable $a$. The last term represents the ambiguity present in the force in general.

\section{Neumann Boundary Conditions}

The calculational procedure to follow in order to compute $\zeta_{\mathscr{M}}$ at $s=-1/2$ for Neumann boundary conditions
closely resembles the one used in the previous section for the Dirichlet case.
We will therefore describe only the few changes that are necessary \cite{bordag96,bordag96a,bordag96b}.
Imposing Neumann boundary conditions in region $I$ leads to the following implicit equation for the eigenvalues $\gamma_{I}$
\begin{equation}\label{45}
\left(\frac{1-d}{2}\right)J_{\nu}(a\gamma_{I})+a\gamma_{I}J'_{\nu}(a\gamma_{I})=0\;,
\end{equation}
while for region $II$ we obtain a linear system of equations for the unknowns $A$ and $B$
\begin{equation}\label{46}
\left\{ \begin{array}{l}
A\left[\left(\frac{1-d}{2}\right)J_{\nu}(a\gamma_{II})+a\gamma_{II}J'_{\nu}(a\gamma_{II})\right]+B\left[\left(\frac{1-d}{2}\right)Y_{\nu}(a\gamma_{II})+a\gamma_{II}Y'_{\nu}(a\gamma_{II})\right]=0\\
A\left[\left(\frac{1-d}{2}\right)J_{\nu}(b\gamma_{II})+b\gamma_{II}J'_{\nu}(b\gamma_{II})\right]+B\left[\left(\frac{1-d}{2}\right)Y_{\nu}(b\gamma_{II})+b\gamma_{II}Y'_{\nu}(b\gamma_{II})\right]=0\; .\\
\end{array} \right.
\end{equation}
This system of equations possesses non-trivial solutions if the following equation holds
\begin{eqnarray}\label{46a}
\lefteqn{\left[\left(\frac{1-d}{2}\right)J_{\nu}(a\gamma_{II})+a\gamma_{II}J'_{\nu}(a\gamma_{II})\right]\left[\left(\frac{1-d}{2}\right)Y_{\nu}(b\gamma_{II})+b\gamma_{II}Y'_{\nu}(b\gamma_{II})\right]}\nonumber\\
&-&\left[\left(\frac{1-d}{2}\right)Y_{\nu}(a\gamma_{II})+a\gamma_{II}Y'_{\nu}(a\gamma_{II})\right]\left[\left(\frac{1-d}{2}\right)J_{\nu}(b\gamma_{II})+b\gamma_{II}J'_{\nu}(b\gamma_{II})\right]=0\;.
\end{eqnarray}
For real $\nu$ and positive $a$, the equation (\ref{45}) possesses only real and simple zeroes if $\nu\geq (d-1)/2a$ \cite{dixon,watson}
and we restrict our attention to this case.

In complete analogy with the Dirichlet case, we obtain the spectral $\zeta$-functions
in the regions $I$ and $II$. To distinguish the results for Neumann boundary conditions from the ones for Dirichlet boundary conditions
we use an upper index $\mathcal{N}$. We find
\begin{equation}
  \zeta^{\mathcal{N}}_{I}(s,a)=\sum_{\nu}d(\nu)\zeta^{\mathcal{N},\,\nu}_{I}(s,a)\;,
\end{equation}
and
\begin{equation}
  \zeta^{\mathcal{N}}_{II}(s,a,b)=\sum_{\nu}d(\nu)\zeta^{\mathcal{N},\,\nu}_{II}(s,a,b)\;,
\end{equation}
with the integral representations valid for $1/2<\Re(s)<1$,
\begin{equation}\label{47}
\zeta^{\mathcal{N},\,\nu}_{I}(s,a)=\frac{\sin(\pi s)}{\pi}\int_{\frac{ma}{\nu}}^{\infty}dk\left[\frac{\nu^{2}k^{2}}{a^2}-m^{2}\right]^{-s}\frac{\partial}{\partial k}\ln \left[k^{-\nu}\left(\beta I_{\nu}(\nu k)+\nu k I'_{\nu}(\nu k)\right)\right]\;,
\end{equation}
and
\begin{equation}\label{48}
\zeta^{\mathcal{N},\,\nu}_{II}(s,a,b)=\frac{\sin(\pi s)}{\pi}\int_{\frac{m}{\nu}}^{\infty}d\kappa\left[(\nu\kappa)^{2}-m^{2}\right]^{-s}\frac{\partial}{\partial \kappa}\ln \Xi_{\nu}(\kappa,a,b)\;.
\end{equation}
In (\ref{48}), for typographical convenience, we have defined $\beta=(1-d)/2$ and
\begin{eqnarray}\label{48a}
\Xi_{\nu}(\kappa,a,b)=\left[\beta I_{\nu}(a\kappa)+a\kappa I'_{\nu}(a\kappa)\right]\left[\beta K_{\nu}(b\kappa)+b\kappa K'_{\nu}(b\kappa)\right]-\left[\beta K_{\nu}(a\kappa)+a\kappa K'_{\nu}(a\kappa)\right]\left[\beta I_{\nu}(b\kappa)+b\kappa I'_{\nu}(b\kappa)\right]\;.\;\;\;\;\;
\end{eqnarray}

In order to perform the analytic continuation of the above integrals to the region $\Re(s)<1/2$ for Neumann boundary conditions we need, in addition to the asymptotic expansion (\ref{17}), the following one for $I^{\prime}_{\nu}(\nu z)$ \cite{gradshtein07,olver54}
\begin{equation}\label{49}
I^{\prime}_{\nu}(\nu z)\sim \frac{1}{\sqrt{{2\pi\nu}}}\frac{e^{\nu\eta}(1+z^{2})^{1/4}}{z}\left[1+\sum_{k=1}^{\infty}\frac{v_{k}(t)}{\nu^{k}}\right]\;,
\end{equation}
where the polynomials $v_{k}(t)$ are determined by the recurrence relation
\begin{equation}\label{50}
v_{k}(t)=u_{k}(t)+t(t^{2}-1)\left[\frac{1}{2}u_{k-1}(t)+t u^{\prime}_{k-1}(t)\right]\;.
\end{equation}
By exploiting (\ref{17}) together with (\ref{49}), we obtain the following uniform asymptotic expansion, which will be useful in the subsequent calculations,
\begin{equation}\label{50a}
\ln\left[\beta I_{\nu}(k\nu)+k\nu I'_{\nu}(k\nu)\right]\sim\ln\left[\sqrt{\frac{\nu}{2\pi}}e^{\nu\eta}(1+k^{2})^{1/4}\right]+\sum_{n=1}^{\infty}\frac{M_{n}(t,\beta)}{\nu^{n}}\;,
\end{equation}
with $M_{n}(t,\beta)$ defined by the cumulant expansion \cite{bordag96,bordag96a,bordag96b,kirsten01}
\begin{equation}\label{51}
\ln\left[1+\sum_{k=1}^{\infty}\frac{v_{k}(t)}{\nu^{k}}+\frac{\beta}{\nu}t\left(1+\sum_{k=1}^{\infty}\frac{u_{k}(t)}{\nu^{k}}\right)\right]\sim\sum_{n=1}^{\infty}\frac{M_{n}(t,\beta)}{\nu^{n}}\; .
\end{equation}
Moreover, the terms $M_{n}(t,\beta)$ have a structure analogous to the $D_{n}(t)$, namely
\begin{equation}\label{52}
M_{n}(t,\beta)=\sum_{i=0}^{n}z_{i,n}(\beta)t^{n+2i}\;,
\end{equation}
where, in this case, the coefficients $z_{i,n}$ depend on the variable $\beta$ (see appendix). With the help of the above uniform asymptotic expansion
we write $\zeta^{\mathcal{N}}_{I}(s)$ as a sum of the terms
\begin{equation}\label{53}
\zeta^{\mathcal{N}}_{I}(s,a)=W_{I}(s,a)+\sum_{i=-1}^{D}A_{i}^{(\mathcal{N},I)}(s,a)\;,
\end{equation}
where one can prove that
\begin{equation}\label{54}
A_{-1}^{(\mathcal{N},I)}(s,a)=A_{-1}^{(I)}(s,a)\quad \mbox{and}\quad A_{0}^{(\mathcal{N},I)}(s,a)=-A_{0}^{(I)}(s,a)\;,
\end{equation}
furthermore for $i\geq 1$, once the coefficients $x_{i,b}$ are replaced with $z_{i,b}$ \cite{bordag96,bordag96a,kirsten01},
\begin{equation}\label{54new}
 A_{i}^{(\mathcal{N},I)}(s,a)=A_{i}^{(I)}(s,a)\;.
\end{equation}
The last term, i.e. $W_{I}(s,a)$, is an analytic function for $-1<\Re(s)<1/2$ defined as
\begin{equation}\label{55}
W_{I}(s,a)=\sum_{\nu}d(\nu)W_{\nu}^{(I)}(s,a)\;,
\end{equation}
with
\begin{eqnarray}\label{56}
W^{(I)}_{\nu}(s,a)&=&a^{2s}\nu^{-2s}\frac{\sin(\pi s)}{\pi}\int_{0}^{\infty}dk\,k^{-2s}\frac{\partial}{\partial k}
\Bigg\{\ln\left[k^{-\nu}(\beta I_{\nu}(k\nu)+k\nu I^{\prime}_{\nu}(k\nu))\right]\nonumber\\
&-&\ln\left[\sqrt{\frac{\nu}{2\pi}}k^{-\nu}e^{\nu\eta}(1+k^{2})^{1/4}\right]
-\sum_{n=1}^{D}\frac{M_{n}(t,\beta)}{\nu^{n}}\Bigg\}\;.\;\;\;\;\;\;
\end{eqnarray}

For the spectral $\zeta$-function $\zeta^{\mathcal{N},\,\nu}_{II}(s,a,b)$ in region $II$ we have a representation similar to the Dirichlet case (\ref{25}),
more explicitly
\begin{eqnarray}\label{57}
\zeta^{\mathcal{N},\,\nu}_{II}(s,a,b)&=&\frac{\sin(\pi s)}{\pi}\int_{\frac{ma}{\nu}}^{\infty}d\kappa\left[\frac{\nu^2\kappa^2}{a^2}-m^{2}\right]^{-s}\frac{\partial}{\partial \kappa}\ln \left[\kappa^{\nu}(-\beta K_{\nu}(\nu\kappa)-\nu\kappa K'_{\nu}(\nu\kappa))\right]\nonumber\\
&+&\frac{\sin(\pi s)}{\pi}\int_{\frac{m b}{\nu}}^{\infty}d\kappa\left[\frac{\nu^{2}\kappa^{2}}{b^{2}}-m^{2}\right]^{-s}\frac{\partial}{\partial \kappa}\ln \left[\kappa^{-\nu}(\beta I_{\nu}(\nu\kappa)+\nu\kappa I'_{\nu}(\nu\kappa))\right]+\mathscr{F}^{\nu}_{\mathcal{N}}(s,a,b)\;,\;\;\;\;\;\;\;\;\;\;\;\;
\end{eqnarray}
where $\mathscr{F}^{\nu}_{\mathcal{N}}(s,a,b)$ is represented by the following integral, once the limit $m\to 0$ is taken,
\begin{equation}\label{58}
\mathscr{F}^{\nu}_{\mathcal{N}}(s,a,b)=\nu^{-2s}\frac{\sin(\pi s)}{\pi}\int_{0}^{\infty}d\kappa\,\kappa^{-2s}\frac{\partial}{\partial \kappa}\ln\Delta_{\nu}(\kappa,a,b)\;,
\end{equation}
with
\begin{equation}\label{59}
\Delta_{\nu}(\kappa,a,b)=1-\frac{\big[\beta I_{\nu}(a\nu\kappa)+a\nu\kappa I'_{\nu}(a\nu\kappa)\big]\big[\beta K_{\nu}(b\nu\kappa)+b\nu\kappa K'_{\nu}(b\nu\kappa)\big]}
{\big[\beta K_{\nu}(a\nu\kappa)+a\nu\kappa K'_{\nu}(a\nu\kappa)\big]\big[\beta I_{\nu}(b\nu\kappa)+b\nu\kappa I'_{\nu}(b\nu\kappa)\big]}\;.
\end{equation}
The domain of analyticity of $\mathscr{F}^{\nu}_{\mathcal{N}}(s,a,b)$ can be found by utilizing arguments analogous to the ones used for $\mathscr{F}^{\nu}_{\mathcal{D}}(s,a,b)$ in the Dirichlet case.
In fact, as $\kappa\to 0$, the integral (\ref{58}) is convergent for $\Re(s)<1/2$. As $\kappa\to\infty$, by using the uniform asymptotic expansions
(\ref{50a}) and (\ref{60a}), we get
\begin{equation}
1-\Delta_{\nu}(\kappa,a,b)\sim\exp\left\{-2\nu\big[\eta(b\kappa)-\eta(a\kappa)\big]\right\}\;,
\end{equation}
which is exponentially suppressed since $\eta(b\kappa)>\eta(a\kappa)$ for $a<b$. Therefore, we can conclude that $\mathscr{F}^{\nu}_{\mathcal{N}}(s,a,b)$ defines an
analytic function for $\Re(s)<1/2$.

The analytic continuation of the integrals in (\ref{57}) is performed exactly like in the Dirichlet case. In particular, we are able to write
$\zeta_{II}^{\mathcal{N}}(s,a,b)$ in the form of a sum
\begin{equation}\label{60}
\zeta_{II}^{\mathcal{N}}(s,a,b)=W_{II}(s,a,b)+\mathscr{F}_{\mathcal{N}}(s,a,b)+\sum_{i=-1}^{D}A_{i}^{(\mathcal{N},II)}(s,a,b)\;,
\end{equation}
where
\begin{equation}
  \mathscr{F}_{\mathcal{N}}(s,a,b)=\sum_{\nu}d(\nu)\mathscr{F}^{\nu}_{\mathcal{N}}(s,a,b)\;,
\end{equation}
and one can prove that
\begin{equation}\label{60b}
A_{i}^{(\mathcal{N},II)}(s,a,b)=(-1)^{i}A_{i}^{(\mathcal{N},I)}(s,a)+A_{i}^{(\mathcal{N},I)}(s,b)\;.
\end{equation}
The representation (\ref{60}) is obtained by exploiting the uniform asymptotic expansion (\ref{50a})
and the following one
\begin{equation}\label{60a}
\ln\left[-\beta K_{\nu}(\nu\kappa)-\nu\kappa K'_{\nu}(\nu\kappa)\right]\sim\ln\left[\sqrt{\frac{\pi\nu}{2}}e^{-\nu\eta}(1+\kappa^{2})^{1/4}\right]+\sum_{n=1}^{\infty}(-1)^{n}\frac{M_{n}(t,\beta)}{\nu^{n}}\;,
\end{equation}
which is obtained from the expansion for $K_{\nu}(\nu\kappa)$ in (\ref{27}) and the one for $K^{\prime}_{\nu}(\nu\kappa)$ \cite{gradshtein07,olver54}
\begin{equation}\label{61}
K'_{\nu}(\nu \kappa)\sim -\sqrt{\frac{\pi}{2\nu}}\frac{(1+\kappa^{2})^{1/4}}{\kappa}e^{-\nu\eta}\left[1+\sum_{n=1}^{\infty}(-1)^{n}\frac{v_{n}(t)}{\nu^{n}}\right]\;.
\end{equation}

The function $W_{II}(s,a,b)$ in (\ref{60}) is analytic in the region $-1<\Re(s)<1/2$ and has the form
\begin{equation}\label{62}
W_{II}(s,a,b)=\sum_{\nu}d(\nu)W_{\nu}^{(II)}(s,a,b)\;,
\end{equation}
where
\begin{eqnarray}\label{63}
W^{(II)}_{\nu}(s,a,b)&=&a^{2s}\nu^{-2s}\frac{\sin(\pi s)}{\pi}\int_{0}^{\infty}d\kappa\,\kappa^{-2s}\frac{\partial}{\partial \kappa}
\Bigg\{\ln\left[-\beta K_{\nu}(\nu\kappa)-\nu\kappa K'_{\nu}(\nu\kappa)\right]
-\ln\left[\sqrt{\frac{\pi\nu}{2}}e^{-\nu\eta}(1+\kappa^{2})^{1/4}\right]\nonumber\\
&-&\sum_{n=1}^{D}(-1)^{n}\frac{M_{n}(t,\beta)}{\nu^{n}}\Bigg\}+b^{2s}\nu^{-2s}\frac{\sin(\pi s)}{\pi}\int_{0}^{\infty}d\kappa\,\kappa^{-2s}\frac{\partial}{\partial \kappa}
\Bigg\{\ln\left[\beta I_{\nu}(\kappa\nu)+\kappa\nu I'_{\nu}(\kappa\nu)\right]\nonumber\\
&-&\ln\left[\sqrt{\frac{\nu}{2\pi}}e^{\nu\eta}(1+\kappa^{2})^{1/4}\right]-\sum_{n=1}^{D}\frac{M_{n}(t,\beta)}{\nu^{n}}\Bigg\}\;.
\end{eqnarray}

The evaluation of the residue and the finite part at $s=-1/2$ of the spectral $\zeta$-function $\zeta_{\mathscr{M}}^{\mathcal{N}}$ for
the conical piston endowed with Neumann boundary conditions proceeds along the same lines of the previous section on
Dirichlet boundary conditions due, in particular, to the relations (\ref{54}), (\ref{54new}) and (\ref{60b}).
For $\zeta_{\mathscr{M}}^{\mathcal{N}}$ in a neighborhood of $s=-1/2$, we obtain the following expression
\begin{eqnarray}\label{64}
\zeta^{\mathcal{N}}_{\mathscr{M}}\left(\alpha-\frac{1}{2},a,b\right)&=&W_{I}\left(-\frac{1}{2},a\right)+W_{II}\left(-\frac{1}{2},a,b\right)+\mathscr{F}_{\mathcal{N}}\left(-\frac{1}{2},a,b\right)\nonumber\\
&+&2\sum_{i=0}^{[D/2]}A^{(\mathcal{N},I)}_{2i}\left(-\frac{1}{2}+\alpha,a\right)+\sum_{i=-1}^{D}A^{(\mathcal{N},I)}_{i}\left(-\frac{1}{2}+\alpha,b\right)\;.
\end{eqnarray}
The residue and finite part of the above spectral $\zeta$-function are readily obtained leading to the result
\begin{eqnarray}\label{65}
\textrm{Res}\,\zeta^{\mathcal{N}}_{\mathscr{M}}\left(-\frac{1}{2},a,b\right)&=&\frac{1}{2}\left(\frac{1}{a}+\frac{1}{2b}\right)\textrm{Res}\,\zeta_{\mathscr{N}}\left(-\frac{1}{2}\right)+\frac{1}{a\sqrt{\pi}}\sum_{i=1}^{[D/2]}\tilde{\omega}_{2i}
\textrm{Res}\,\zeta_{\mathscr{N}}\left(\frac{2i-1}{2}\right)-\frac{1}{4\pi b}\zeta_{\mathscr{N}}\left(-1\right)\nonumber\\
&-&\frac{1}{2\pi b}\left(\frac{3}{8}-\beta\right)\zeta_{\mathscr{N}}\left(0\right)+\frac{1}{2\sqrt{\pi}b}\sum_{i=2}^{D}\tilde{\omega}_{i}\textrm{Res}\,\zeta_{\mathscr{N}}\left(\frac{i-1}{2}\right)\;,
\end{eqnarray}
and, for the finite part,
\begin{eqnarray}\label{66}
\textrm{FP}\,\zeta^{\mathcal{N}}_{\mathscr{M}}\left(-\frac{1}{2},a,b\right)&=&W_{I}\left(-\frac{1}{2},a\right)+W_{II}\left(-\frac{1}{2},a,b\right)+\mathscr{F}_{\mathcal{N}}\left(-\frac{1}{2},a,b\right)+\frac{1}{2a}\left[\textrm{FP}\,\zeta_{\mathscr{N}}\left(-\frac{1}{2}\right)+\ln a^{2}\,\textrm{Res}\,\zeta_{\mathscr{N}}\left(-\frac{1}{2}\right)\right]\nonumber\\
&+&\frac{1}{a\sqrt{\pi}}\sum_{i=1}^{[D/2]}\Bigg[\tilde{\omega}_{2i}\textrm{FP}\,\zeta_{\mathscr{N}}\left(\frac{2i-1}{2}\right)
+\tilde{\omega}_{2i}\left(\ln a^{2}+\gamma+2\ln 2-2\right)\textrm{Res}\,\zeta_{\mathscr{N}}\left(\frac{2i-1}{2}\right)\nonumber\\
&+&\tilde{\Omega}_{2i}\textrm{Res}\,\zeta_{\mathscr{N}}\left(\frac{2i-1}{2}\right)\Bigg]-\frac{1}{4\pi b}\left[\Big(2\ln 2+1\Big)\zeta_{\mathscr{N}}(-1)+\zeta'_{\mathscr{N}}(-1)\right]+\frac{1}{4b}\textrm{FP}\,\zeta_{\mathscr{N}}\left(-\frac{1}{2}\right)\nonumber\\
&-&\frac{1}{2\pi b}\left(\frac{3}{8}-\beta\right)\left[\zeta'_{\mathscr{N}}(0)+\zeta_{\mathscr{N}}(0)\left(2\ln 2-2\right)\right]-\frac{7}{24\pi b}\zeta_{\mathscr{N}}(0)+\frac{1}{2\sqrt{\pi}b}\sum_{i=2}^{D}\Bigg[\tilde{\omega}_{i}\textrm{FP}\,\zeta_{\mathscr{N}}\left(\frac{i-1}{2}\right)\nonumber\\
&+&\tilde{\omega}_{i}\left(\gamma+2\ln 2-2\right)\textrm{Res}\,\zeta_{\mathscr{N}}\left(\frac{i-1}{2}\right)+\tilde{\Omega}_{i}\textrm{Res}\,\zeta_{\mathscr{N}}\left(\frac{i-1}{2}\right)\Bigg]\;,
\end{eqnarray}
where, in the above formulas, we have introduced the notation
\begin{equation}\label{67}
\tilde{\omega}_{i}=\sum_{p=0}^{i}z_{i,p}\frac{\Gamma\left(p+\frac{i-1}{2}\right)}{\Gamma\left(p+\frac{i}{2}\right)}\;,\qquad
\tilde{\Omega}_{i}=\sum_{p=0}^{i}z_{i,p}\frac{\Gamma\left(p+\frac{i-1}{2}\right)}{\Gamma\left(p+\frac{i}{2}\right)}\Psi\left(p+\frac{i-1}{2}\right)\; ,
\end{equation}
which is identical to (\ref{40}) once $x_{i,p}$ is replaced by $z_{i,p}$.

By exploiting the relation (\ref{10b}) we obtain the expression for the force on the piston
when Neumann boundary conditions are imposed
\begin{eqnarray}\label{68}
F^{\textrm{Neu}}_{\textrm{Cas}}(a,b)&=&-\frac{1}{2}W'_{I}\left(-\frac{1}{2},a\right)-\frac{1}{2}W'_{II}\left(-\frac{1}{2},a\right)-\frac{1}{2}\mathscr{F}'_{\mathcal{N}}\left(-\frac{1}{2},a,b\right)-\frac{1}{4a^{2}}\left[(2-\ln a^{2})\textrm{Res}\,\zeta_{\mathscr{N}}\left(-\frac{1}{2}\right)-\textrm{FP}\,\zeta_{\mathscr{N}}\left(-\frac{1}{2}\right)\right]\nonumber\\
&+&\frac{1}{2\sqrt{\pi}a^{2}}\sum_{i=1}^{[D/2]}\Bigg[\tilde{\omega}_{2i}\textrm{FP}\,\zeta_{\mathscr{N}}\left(\frac{2i-1}{2}\right)
-\tilde{\omega}_{2i}\left(4-\ln a^{2}-\gamma-2\ln 2\right)\textrm{Res}\,\zeta_{\mathscr{N}}\left(\frac{2i-1}{2}\right)\nonumber\\
&+&\tilde{\Omega}_{2i}\textrm{Res}\,\zeta_{\mathscr{N}}\left(\frac{2i-1}{2}\right)\Bigg]
+\frac{1}{4a^{2}}\left(\frac{1}{\alpha}+\ln\mu^{2}\right)\Bigg[
\textrm{Res}\,\zeta_{\mathscr{N}}\left(-\frac{1}{2}\right)+\frac{2}{\sqrt{\pi}}\sum_{i=1}^{[D/2]}\tilde{\omega}_{2i}\textrm{Res}\,\zeta_{\mathscr{N}}\left(\frac{2i-1}{2}\right)\Bigg]+O(\alpha)\;.\nonumber\\
\end{eqnarray}
The expressions (\ref{44}) and (\ref{68}) represent the Casimir force on the piston, endowed, respectively, with Dirichlet and Neumann boundary conditions, which is valid in any dimension
and for any compact and smooth manifold $\mathscr{N}$. It is important to stress that more explicit results can be obtained once
the geometry of the manifold $\mathscr{N}$ has been specified. In addition, the analytic functions that appear in the results (\ref{44}) and (\ref{68}), namely
$Z_{I}$, $Z_{II}$, $W_{I}$, $W_{II}$, $\mathscr{F}_{\mathcal{D}}$, and $\mathscr{F}_{\mathcal{N}}$ can only be handled in
a numerical way. Although the expressions for the force are written in terms of an infinite series in the eigenvalues $\nu$, in practice, one only sums finitely
many terms in such a way to obtain a numerical result with a prescribed accuracy.

These results also show explicitly the terms that are responsible for the ambiguity in the Casimir force. A closer look at these terms
demonstrates that they are proportional to the heat kernel coefficients $a_{(d+1)/2-i}$ of the manifold $\mathscr{N}$ with $0\leq i\leq [(d+1)/2]$.
It is clear that this ambiguity in the prediction of the force disappears if the manifold $\mathscr{N}$ is even-dimensional without boundary. However, even if a suitable manifold is chosen for which the Casimir force on the piston is well defined, the formulas  (\ref{44}) and (\ref{68})
do not allow to extract any information about its sign analytically, instead a numerical study is necessary; see Section VI.

In the next section we will study the asymptotic behavior of the Casimir force on the piston when both $a$ and $b$ are large, and when $a\to 0$, i.e. the piston approaches the conical singularity.

\section{Asymptotic Behavior of the Casimir force for both $a$ and $b$ large and for $a\to 0$}

In this section we will study the asymptotic behavior of the Casimir force for Dirichlet and Neumann boundary conditions when both
the parameters $a$ and $b$ are large, namely $b/a\to 1$. For this analysis, we will employ the arguments outlined in \cite{teo10} for the case of concentric spheres. Moreover, we will describe in detail only the Dirichlet case, since the Neumann case can be treated in exactly the same way. In the limit $b/a\to 1$ the Casimir force for the conical piston will reproduce the
one for parallel plates. From the expressions (\ref{44}) and (\ref{68}) it is not difficult to realize that all the terms but $\mathscr{F}_{\mathcal{D}}$ and
$\mathscr{F}_{\mathcal{N}}$ vanish proportionally to $a^{-2}$ as $a\to \infty$. Therefore, the asymptotic behavior for $a$ and $b$ large when Dirichlet or Neumann boundary conditions are imposed is given, respectively, by $\mathscr{F}_{\mathcal{D}}$ and $\mathscr{F}_{\mathcal{N}}$. With the last remark in mind we can write that, when $b/a\to 1$,
\begin{equation}\label{wil}
F^{\textrm{Dir}}_{\textrm{Cas}}(a,b)\sim -\frac{1}{2\pi}\sum_{\nu}d(\nu)\nu\int_{0}^{\infty}d\kappa\frac{\partial}{\partial a}\left[1-\frac{K_{\nu}(\nu\kappa b)I_{\nu}(\nu\kappa a)}{K_{\nu}(\nu\kappa a)I_{\nu}(\nu\kappa b)}\right]\;.
\end{equation}
The derivative in the integrand, denoted by $\mathcal{P} (\kappa , a,b)$, can be evaluated my making use of the properties of the derivative of the modified Bessel functions to get
\begin{equation}\label{wil1}
\mathcal{P}(\kappa,a,b)=-\frac{K_{\nu}(\nu\kappa b)}{a K^{2}_{\nu}(\nu\kappa a)I_{\nu}(\nu\kappa b)}\left[1-\frac{K_{\nu}(\nu\kappa b)I_{\nu}(\nu\kappa a)}{K_{\nu}(\nu\kappa a)I_{\nu}(\nu\kappa b)}\right]^{-1}\;.
\end{equation}
By performing the following change of variables $\kappa\to\kappa/a$, we obtain the expression
\begin{equation}\label{c}
F^{\textrm{Dir}}_{\textrm{Cas}}(a,b)\sim \frac{1}{2\pi a^{2}}\sum_{\nu}d(\nu)\nu\int_{0}^{\infty}d\kappa \,\mathcal{P}\left(\frac{\nu\kappa}{a},1,\frac{b}{a}\right)\;.
\end{equation}
The $b\to a$ behavior of the Casimir force is already captured using the uniform asymptotic expansion of the modified Bessel
functions to obtain \cite{teo10}
\begin{equation}\label{a}
\mathcal{P}\left(\frac{\nu\kappa}{a},1,\frac{b}{a}\right)\sim 2\nu\sqrt{1+\kappa^{2}}\sum_{n=1}^{\infty}e^{-2n\nu\left[\eta\left(\frac{b\kappa}{a}\right)-\eta(\kappa)\right]}
\sum_{i=0}^{\infty}\frac{p_{i}(t,a,b,n)}{\nu^{i}}\;,
\end{equation}
where $p_{i}$ are polynomials in $t$ which vanish when $b\to a$ and such that $p_{0}=1$ \cite{teo10}. By utilizing the inverse Mellin transform we
can rewrite the sum of the exponentials in (\ref{a}) in terms of a complex integral as follows
\begin{equation}\label{b}
\sum_{n=1}^{\infty}e^{-2n\nu\left[\eta\left(\frac{b\kappa}{a}\right)-\eta(\kappa)\right]}=\sum_{n=1}^{\infty}\frac{1}{2\pi i}\int_{c-i\infty}^{c+i\infty}d\alpha\,\Gamma(\alpha)(2n\nu)^{-\alpha}\left[\eta\left(\frac{b\kappa}{a}\right)-\eta(\kappa)\right]^{-\alpha}\;,
\end{equation}
which is valid for $\Re(\nu)>0$, and we assume that $\Re (c)$ is large enough so that the sum over $\nu$ and the integral in $\alpha$ can be safely interchanged.

By substituting the expressions (\ref{a}) and (\ref{b}) into the integral (\ref{c}), and by recalling the definition of $\zeta_{\mathscr{N}}$ in (\ref{10})
we obtain
\begin{eqnarray}\label{d}
F^{\textrm{Dir}}_{\textrm{Cas}}(a,b)\sim\frac{1}{2\pi^{2}i\,a^{2}}\sum_{i=0}^{\infty}\int_{c-i\infty}^{c+i\infty}d\alpha\,\Gamma(\alpha)(2)^{-\alpha}\zeta_{R}(\alpha)\zeta_{\mathscr{N}}\left(\frac{\alpha+i}{2}-1\right)\nonumber\\
\times\int_{0}^{\infty}d\kappa\sqrt{1+\kappa^{2}}\left[\eta\left(\frac{b\kappa}{a}\right)-\eta(\kappa)\right]^{-\alpha}p_{i}(t,a,b,n)\;,
\end{eqnarray}
where the integral over $\kappa$ is convergent for $\Re (c)>2$, since $\eta\left(b\kappa/a\right)-\eta(\kappa)\to \ln(b/a)$ as $\kappa\to 0$ and $\eta\left(b\kappa/a\right)-\eta(\kappa)\to(b/a-1)\kappa$ as $\kappa\to\infty$ \cite{teo10}.

It is convenient, at this point, to introduce the variable $q=b/a-1$. In this new variable the limit we are interested in corresponds to $q\to 0$.
The leading behavior in $q$ of the integral (\ref{d}) is given by the term with $i=0$, since the other polynomials $p_{i}$, with $i\geq 1$, vanish as $q\to 0$, more explicitly
\begin{equation}\label{e}
F^{\textrm{Dir}}_{\textrm{Cas}}(q)\sim\frac{1}{2\pi^{2}i\,a^{2}}\int_{c-i\infty}^{c+i\infty}d\alpha\,\Gamma(\alpha)(2)^{-\alpha}\zeta_{R}(\alpha)\zeta_{\mathscr{N}}\left(\frac{\alpha}{2}-1\right)\int_{0}^{\infty}d\kappa\sqrt{1+\kappa^{2}}\left[\eta\left((q+1)\kappa\right)-\eta(\kappa)\right]^{-\alpha}\;.
\end{equation}
By closing the contour to the right, we encounter all poles of $\zeta_{\mathscr{N}}$ with $\Re(\alpha)>2$.
The rightmost pole of $\zeta_{\mathscr{N}}$ results from $\alpha = d+2$, the other poles further to the left will be irrelevant for what follows because
the leading behavior of (\ref{e}) as $q\to 0$ is proportional to the residue of $\zeta_{\mathscr{N}} (\alpha /2-1)$ at $\alpha=d+2$. This is seen by using \cite{teo10}
\begin{equation}\label{f}
\eta\left((q+1)\kappa\right)-\eta(\kappa)=q\sqrt{1+\kappa^{2}}+O(q^{2})\;.
\end{equation}

The last remark, and the expansion (\ref{f}), allow us to obtain the Casimir force for the conical piston endowed with Dirichlet boundary conditions when $b/a\to 1$
as follows
\begin{equation}\label{h}
F^{\textrm{Dir}}_{\textrm{Cas}}(q)=\frac{\Gamma(D+1)\zeta_{R}(D+1)}{2^{D+1}\sqrt{\pi}\,\Gamma\left(\frac{D}{2}\right)}\frac{\mathscr{A}^{\mathscr{N}}_{0}}{q^{D+1}}+O\left(q^{-D}\right)\;,
\end{equation}
where the higher orders in $q$ would involve the other heat kernel coefficients of $\mathscr{N}$. The Neumann case is treated in exactly the same way and the resulting Casimir force when $b/a\to 1$ reduces to (\ref{h}).

Let us now turn our attention to the limiting behavior of the Casimir force when the piston approaches the conical singularity, namely, when $a\to 0$.
From the expressions (\ref{44}) and (\ref{68}) it is not difficult to observe that all the terms, except $\mathscr{F}_{\mathcal{D}}$ and $\mathscr{F}_{\mathcal{N}}$,
are proportional to $a^{-2}$. The proportionality coefficient and its sign will depend on the specific geometry of the piston $\mathscr{N}$. It is therefore sufficient to study
the asymptotics of $\mathscr{F}_{\mathcal{D}}$ and $\mathscr{F}_{\mathcal{N}}$ as $a\to 0$. Once again, we will focus on the Dirichlet case since the Neumann case can be treated in exactly the same way.

From the expressions (\ref{26}) and (\ref{wil1}), by exploiting the change of variables $\kappa\to\kappa/\nu$, we can write
\begin{equation}
-\frac{1}{2}\mathscr{F}'_{\mathcal{D}}\left(-\frac{1}{2},a,b\right)=-\frac{1}{2\pi}\sum_{\nu}d(\nu)\nu\int_{0}^{\infty}d\kappa \,\mathcal{P}\left(\frac{\kappa}{\nu},a,b\right)\;.
\end{equation}
Since we are interested in the small $a$ expansion, we utilize the series representation of the modified Bessel functions to obtain the following
expression, to leading order in $a$, of (\ref{wil1})
\begin{equation}
\mathcal{P}\left(\frac{\kappa}{\nu},a,b\right)=\frac{4}{\Gamma^{2}(\nu)}a^{2\nu-1}\frac{K_{\nu}(\kappa b)}{I_{\nu}(\kappa b)}\left(\frac{\kappa}{2}\right)^{2\nu}+O\left(\kappa^{2\nu+2}\right)\;.
\end{equation}
This expansion allows us to write
\begin{equation}
-\frac{1}{2}\mathscr{F}'_{\mathcal{D}}\left(-\frac{1}{2},a,b\right)=\frac{2}{\pi}\sum_{\nu}d(\nu)\frac{a^{2\nu-1}}{\Gamma^{2}(\nu)}\int_{0}^{\infty}d\kappa \frac{K_{\nu}(\kappa b)}{I_{\nu}(\kappa b)}\left(\frac{\kappa}{2}\right)^{2\nu}+O\left(\kappa^{2\nu+2}\right)\;.
\end{equation}
The integral in the last expression is convergent, since
\begin{equation}
\frac{K_{\nu}(\kappa b)}{I_{\nu}(\kappa b)}\left(\frac{\kappa}{2}\right)^{2\nu}\sim \frac{1}{2}\Gamma(\nu)\Gamma(\nu+1)b^{-2\nu}\;,\quad\textrm{for}\quad \kappa\to 0\;,
\end{equation}
and
\begin{equation}
\frac{K_{\nu}(\kappa b)}{I_{\nu}(\kappa b)}\left(\frac{\kappa}{2}\right)^{2\nu}\sim \pi e^{-2\kappa}\left(\frac{\kappa}{2}\right)^{2\nu}\;,\quad\textrm{for}\quad \kappa\to \infty\;.
\end{equation}
We can therefore conclude that for $\nu > -1/2$, which is within the assumptions of our work, the contribution $\mathscr{F}'_{\mathcal{D}}$ is subleading as $a\to 0$.

The previous arguments show that as $a\to 0$ the Casimir force on the piston possesses a behavior of the type $a^{-2}$ with the proportionality coefficient depending
on $\zeta_{\mathscr{N}}$ and, hence, on the geometry of the piston.

In the next section we will study a particular case in which the manifold $\mathscr{N}$ is a $d$-dimensional sphere.
This case is of special interest because the spectral zeta function of a $d$-dimensional sphere can be explicitly
evaluated in terms of a linear combination of Hurwitz $\zeta$-functions \cite{bordag96,chang,kirsten01}.

\section{A Specific Piston: The $d$-dimensional Sphere}

In this section we will study the case in which the base manifold is a $d$-dimensional sphere. In this particular
situation the eigenvalues of the Laplacian on ${\mathscr{N}}$ are known to be
\begin{equation}\label{69}
\nu=\left(l+\frac{d-1}{2}\right)\;,
\end{equation}
with $l\geq 0$, and the eigenfunctions are hyperspherical harmonics with degeneracy
\begin{equation}\label{70}
d(l)=(2l+d-1)\frac{(l+d-2)!}{l!(d-1)!}\;.
\end{equation}
The explicit knowledge of the eigenvalues $\nu$ and their degeneracy allows us
to write the $\zeta$-function on ${\mathscr{N}}$ as follows
\begin{equation}\label{71}
\zeta_{{\mathscr{N}}}(s)=\sum_{l=0}^{\infty}(2l+d-1)\frac{(l+d-2)!}{l!(d-1)!}\left(l+\frac{d-1}{2}\right)^{-2s}\;.
\end{equation}
The previous expression can be conveniently rewritten in terms of a linear combination of Hurwitz $\zeta$-functions, namely
\cite{bordag96,bordag96a}
\begin{equation}\label{72}
\zeta_{{\mathscr{N}}}(s)=2\sum_{\alpha=0}^{d-1}e_{\alpha}\zeta_{H}\left(2s-\alpha-1,\frac{d-1}{2}\right)\;,
\end{equation}
where the coefficients $e_{\alpha}$ are defined through the relation
\begin{equation}\label{73}
\frac{(l+d-2)!}{l!(d-1)!}=\sum_{\alpha=0}^{d-1}e_{\alpha}\left(l+\frac{d-1}{2}\right)^{\alpha}\;.
\end{equation}
As one can easily see, from the general results (\ref{44}) and (\ref{68}), we need to compute
either the value or the residue or the finite part of the spectral $\zeta$-function on the manifold $\mathscr{N}$
at specific points. For $s=-m/2$, with $m\geq -1$, by utilizing the equation (\ref{72}), we obtain the result \cite{barnes03a,barnes03b,chang,dowker94}
\begin{equation}\label{74}
\zeta_{\mathscr{N}}\left(-\frac{m}{2}\right)=-2\sum_{\alpha=0}^{d-1}\frac{e_{\alpha}}{m+\alpha+2}B_{m+\alpha+2}\left(\frac{d-1}{2}\right)\;,
\end{equation}
where we have used the relation, valid for $n\geq 0$ \cite{gradshtein07},
\begin{equation}\label{75}
\zeta_{H}(-n,q)=-\frac{B_{n+1}(q)}{n+1}\;,
\end{equation}
with $B_{n}(q)$ being the Bernoulli polynomials. It is clear from the relation (\ref{74}) that the residue of
$\zeta_{\mathscr{N}}(s)$ at $s=-1/2$ and $s=1/2$ vanish.

The remaining values of $s$ that we need to analyze are the ones of the form $s=m/2$ with $m\geq 0$. It is well known
that the Hurwitz $\zeta$-function, $\zeta_{H}(s,q)$, has a simple pole at $s=1$. The explicit form of its Laurent expansion depends on whether
$q$ is an integer or a half-integer. In fact, for $q=n+1/2$, with $n\geq 0$, we have
\begin{equation}\label{76}
\zeta_{H}\left(1+\varepsilon x,n+\frac{1}{2}\right)=\frac{1}{\varepsilon x}+\gamma+2\ln 2-2\sum_{k=1}^{n}\frac{1}{2k-1}+O(x)\;,
\end{equation}
while for $q=n$ and $n\geq 0$ we obtain
\begin{equation}\label{77}
\zeta_{H}\left(1+\varepsilon x,n\right)=\frac{1}{\varepsilon x}+\gamma-\sum_{k=1}^{n-1}\frac{1}{k}+O(x)\;.
\end{equation}
By utilizing the above results and the relation (\ref{72}) we obtain the following expression
for the residue of $\zeta_{\mathscr{N}}(s)$ at $s=m/2$ with $d\geq m\geq 2$,
\begin{equation}\label{78}
\textrm{Res}\,\zeta_{\mathscr{N}}\left(\frac{m}{2}\right)=e_{m-2}\;,
\end{equation}
the contribution to the residue coming from
the index $\alpha = m-2$ of the summation in (\ref{72}). According to (\ref{76}) and (\ref{77}), for the finite part at $s=m/2$ we have to distinguish between two cases. If the dimension $d$ of the manifold
$\mathscr{N}$ is {\it even}, then
\begin{equation}\label{79}
\textrm{FP}\,\zeta_{\mathscr{N}}\left(\frac{m}{2}\right)=2\sum_{{\alpha=0 \atop \alpha\neq m-2}}^{d-1}e_{\alpha}\zeta_{H}\left(m-\alpha-1,\frac{d-1}{2}\right)+2e_{m-2}\left(\gamma+2\ln 2-2\sum_{k=1}^{\frac{d}{2}-1}\frac{1}{2k-1}\right)\;,
\end{equation}
while if $d$ is {\it odd}, we have
\begin{equation}\label{80}
\textrm{FP}\,\zeta_{\mathscr{N}}\left(\frac{m}{2}\right)=2\sum_{{\alpha=0 \atop \alpha\neq m-2}}^{d-1}e_{\alpha}\zeta_{H}\left(m-\alpha-1,\frac{d-1}{2}\right)+2e_{m-2}\left(\gamma-\sum_{k=1}^{\frac{d-3}{2}}\frac{1}{2k-1}\right)\;.
\end{equation}

The results that we have obtained in this section so far are well suited for the analysis of conical pistons with Dirichlet boundary conditions.
However, the case of Neumann boundary conditions needs a special treatment when the manifold $\mathscr{N}$ is a $d$-dimensional sphere.
In fact, from the integral representation (\ref{47}) of the spectral $\zeta$-function in region $I$, one can notice that the case $l=0$, namely
$\nu=(d-1)/2$, needs particular care because the behavior of $\left(\beta I_{\frac{d-1}{2}}(a k)+a k I'_{\frac{d-1}{2}}(a k)\right)$ as $k\to 0$ is different.
This corresponds, according to (\ref{69}), to the lowest angular eigenvalue $\nu$.
In this case, in fact, we have the following small $k$ expansion
\begin{equation}\label{81}
\left(\beta I_{\frac{d-1}{2}}(a k)+a k I'_{\frac{d-1}{2}}(a k)\right)=\frac{(ak)^{\frac{d+3}{2}}}{2^{\frac{d+1}{2}}\Gamma\left(\frac{d+3}{2}\right)}+O\left(k^{\frac{d+7}{2}}\right)\;.
\end{equation}
The above expansion suggests that a more suitable integral representation for the contribution of the mode $\nu = (d-1)/2$
is the following
\begin{equation}\label{82}
\zeta^{\mathcal{N},\,l=0}_{I}(s,a)=a^{2s}\left(\frac{d-1}{2}\right)\frac{\sin(\pi s)}{\pi}\int_{0}^{\infty}dk\,k^{-2s}\frac{\partial}{\partial k}\ln \left[k^{-\frac{d+3}{2}}\left(\beta I_{\frac{d-1}{2}}( k)+ k I'_{\frac{d-1}{2}}( k)\right)\right]\;,
\end{equation}
which is valid in the strip $1/2< \Re(s)< 1$. In order to analytically continue the above expression in the neighborhood of $s=-1/2$, we utilize the following $k\to\infty$ asymptotic expansion
\begin{equation}\label{83}
\ln \left[k^{-\frac{d+3}{2}}\left(\beta I_{\frac{d-1}{2}}( k)+ k I'_{\frac{d-1}{2}}( k)\right)\right]\sim \ln\left[\frac{k^{-\frac{d}{2}-1}}{\sqrt{2\pi}}\,e^{k}\right]
+\sum_{n=1}^{\infty}\frac{\mathcal{B}_{n}\left(\frac{d-1}{2}\right)}{k^{n}}\;,
\end{equation}
where the polynomials are defined according to the relation
\begin{equation}\label{84}
\left[1+\sum_{n=1}^{\infty}(-1)^{n}\frac{r_{n}\left(\frac{d-1}{2}\right)}{k^{n}}+\frac{1-d}{2k}\left(1+\sum_{n=1}^{\infty}(-1)^{n}\frac{p_{n}\left(\frac{d-1}{2}\right)}{k^{n}}\right)\right]
\sim\sum_{n=1}^{\infty}\frac{\mathcal{B}_{n}\left(\frac{d-1}{2}\right)}{k^{n}}\;.
\end{equation}
The expressions (\ref{83}) and (\ref{84}) can be obtained by exploiting the asymptotic expansions for $z\to\infty$ \cite{olver},
\begin{equation}
I_{\nu}(z)\sim \frac{e^{z}}{\sqrt{2\pi z}}\left[1+\sum_{n=1}^{\infty}\frac{r_{n}(\nu)}{z^{n}}\right]\;
\quad\textrm{and}\quad I'_{\nu}(z)\sim \frac{e^{z}}{\sqrt{2\pi z}}\left[1+\sum_{n=1}^{\infty}\frac{p_{n}(\nu)}{z^{n}}\right]\;,
\end{equation}
where the functions $r_{n}(x)$ and $p_{n}(x)$ are obtained from the relations
\begin{equation}\label{85}
r_{0}(x)=1\;,\quad r_{1}(x)=\frac{4x^2+3}{8}\;,\quad r_{n}(x)=\frac{4x^{2}+4n^{2}-1}{n!8^{n}}\prod_{i=2}^{n}\left[4x^2-(2i-3)^2\right]\;,
\end{equation}
and
\begin{equation}\label{86}
p_{0}(x)=1\;,\quad p_{n}(x)=\frac{1}{n!8^{n}}\prod_{i=1}^{n}\left[4x^2-(2i-1)^2\right]\;.
\end{equation}
At this point it will be sufficient, for our purposes, to add and subtract the leading term of the expansion to obtain the result which
is valid for $-1<\Re(s)<1/2$,
\begin{eqnarray}\label{87}
\zeta^{\mathcal{N},\,l=0}_{I}(s,a)&=&W^{l=0}_{I}(s,a)-a^{2s}\frac{\sin(\pi s)}{\pi}\left(\frac{d-1}{2}\right)\Bigg\{\left(\frac{d}{2}+1\right)\frac{1}{2s}
-\frac{1}{2s-1}\nonumber\\
&-&\frac{1}{2s+1}\left(\frac{3}{8}+\frac{d-1}{2}+\frac{(d-1)^{2}}{8}\right)\Bigg\}\;,
\end{eqnarray}
where
\begin{eqnarray}
W^{l=0}_{I}(s,a)&=&a^{2s}\left(\frac{d-1}{2}\right)\frac{\sin(\pi s)}{\pi}\int_{0}^{\infty}dk\,k^{-2s}\frac{\partial}{\partial k}\Bigg\{\ln \left[k^{-\frac{d+3}{2}}\left(\beta I_{\frac{d-1}{2}}( k)+ k I'_{\frac{d-1}{2}}( k)\right)\right]\nonumber\\
&-&H(k-1)\Bigg[\ln\left(\frac{k^{-\frac{d}{2}-1}}{\sqrt{2\pi}}\,e^{k}\right)-\frac{1}{k}\left(\frac{3}{8}+\frac{d-1}{2}+\frac{(d-1)^{2}}{8}\right)\Bigg]\Bigg\}\;,
\end{eqnarray}
with the step function $H(k)$.
In region $II$ a similar argument for the lowest angular eigenvalue is necessary. In fact by utilizing the small $\kappa$ expansion
of the modified Bessel functions, it is not very difficult to show that $\Xi_{\frac{d-1}{2}}(a,\kappa)$ in (\ref{48a}) behaves differently as $\kappa\to 0$. More specifically, one obtains
\begin{equation}
\Xi_{\frac{d-1}{2}}(a,\kappa)=\frac{2a^{2}}{d+3}\left[\left(\frac{a}{b}\right)^{-\frac{d+3}{2}}-\left(\frac{a}{b}\right)^{\frac{d-1}{2}}\right]\left(\frac{\kappa}{2}\right)^{2}+O\left(\frac{\kappa^{4}}{16}\right)\;.
\end{equation}
A suitable representation
for the spectral $\zeta$-function in region $II$ corresponding to the lowest eigenvalue is therefore
\begin{eqnarray}\label{88}
\zeta^{\mathcal{N},\,l=0}_{II}(s,a)&=&a^{2s}\left(\frac{d-1}{2}\right)\frac{\sin(\pi s)}{\pi}\int_{0}^{\infty}d\kappa\,\kappa^{-2s}\frac{\partial}{\partial \kappa}\ln \left[\kappa^{\frac{d-1}{2}}(-\beta K_{\frac{d-1}{2}}(\kappa)-\kappa K'_{\frac{d-1}{2}}(\kappa))\right]\nonumber\\
&+&b^{2s}\left(\frac{d-1}{2}\right)\frac{\sin(\pi s)}{\pi}\int_{0}^{\infty}d\kappa\,\kappa^{-2s}\frac{\partial}{\partial \kappa}\ln \left[\kappa^{-\frac{d+3}{2}}(\beta I_{\frac{d-1}{2}}(\kappa)+\kappa I'_{\frac{d-1}{2}}(\kappa))\right]\nonumber\\
&+&\left(\frac{d-1}{2}\right)\frac{\sin(\pi s)}{\pi}\int_{0}^{\infty}d\kappa\,\kappa^{-2s}\frac{\partial}{\partial \kappa}\ln\Delta_{\frac{d-1}{2}}(\kappa,a,b)\;.
\end{eqnarray}
After analytic continuation of (\ref{88}), we obtain the following result which
is well defined in the region $-1<\Re(s)<1/2$
\begin{eqnarray}\label{89}
\zeta^{\mathcal{N},\,l=0}_{II}(s,a,b)&=&W^{l=0}_{II}(s,a,b)+\mathscr{F}_{\mathcal{N}}^{l=0}(s,a,b)
-\frac{\sin(\pi s)}{\pi}\left(\frac{d-1}{2}\right)\Bigg\{\frac{1}{2s}\left[b^{2s}-\frac{d}{2}\left(a^{2s}-b^{2s}\right)\right]
+\frac{a^{2s}-b^{2s}}{2s-1}\nonumber\\
&+&\frac{1}{2s+1}\left(\frac{3}{8}+\frac{d-1}{2}+\frac{(d-1)^{2}}{8}\right)\left(a^{2s}-b^{2s}\right)\Bigg\}\;,
\end{eqnarray}
where $W^{l=0}_{II}(s,a,b)$ has the form
\begin{eqnarray}\label{90}
W^{l=0}_{II}(s,a,b)&=&\left(\frac{d-1}{2}\right)\frac{\sin(\pi s)}{\pi}\int_{0}^{\infty}d\kappa\,\kappa^{-2s}\frac{\partial}{\partial \kappa}\Bigg\{a^{2s}\ln \left[\kappa^{\frac{d-1}{2}}\left(-\beta K_{\frac{d-1}{2}}( \kappa)- \kappa K'_{\frac{d-1}{2}}( \kappa)\right)\right]\nonumber\\
&+&b^{2s}\ln \left[\kappa^{-\frac{d+3}{2}}\left(\beta I_{\frac{d-1}{2}}( \kappa)+ \kappa I'_{\frac{d-1}{2}}( \kappa)\right)\right]-H(\kappa-1)\Bigg[a^{2s}\ln\left(\sqrt{\frac{\pi}{2}}\,\kappa^{\frac{d}{2}}e^{-\kappa}\right)+b^{2s}\ln\left(\frac{\kappa^{-\frac{d}{2}-1}}{\sqrt{2\pi}}\,e^{\kappa}\right)\nonumber\\
&+&\frac{1}{\kappa}\left(\frac{3}{8}+\frac{d-1}{2}+\frac{(d-1)^{2}}{8}\right)\left(a^{2s}-b^{2s}\right)\Bigg]\Bigg\}\;.
\end{eqnarray}
Let us mention that $\mathscr{F}_{\mathcal{N}}^{l=0}(s,a,b)$ coincides with (\ref{58}) once we set $\nu=(d-1)/2$.

From the above results, it is not difficult to obtain the residue and finite part for $\zeta_{\mathscr{M}}^{\mathcal{N},\,l=0}$ at $s=-1/2$ corresponding to the lowest angular eigenvalue.
In more detail, we have
\begin{equation}\label{91}
\textrm{FP}\,\zeta_{\mathscr{M}}^{\mathcal{N},\,l=0}\left(-\frac{1}{2},a,b\right)=W^{l=0}_{I}\left(-\frac{1}{2},a\right)+W^{l=0}_{II}\left(-\frac{1}{2},a,b\right)+\mathscr{F}_{\mathcal{N}}^{l=0}\left(-\frac{1}{2},a,b\right)
-\frac{d-1}{2\pi}\left(\frac{1}{a}+\frac{d+1}{2b}\right)\;,
\end{equation}
and
\begin{equation}\label{92}
\textrm{Res}\,\zeta_{\mathscr{M}}^{\mathcal{N},\,l=0}\left(-\frac{1}{2},a\right)=-\frac{d-1}{4\pi b}\left(\frac{3}{8}+\frac{d-1}{2}+\frac{(d-1)^{2}}{8}\right)\;.
\end{equation}
Therefore, the contribution of $\nu=(d-1)/2$ to the Casimir force on the piston when Neumann boundary conditions are imposed is, according to (\ref{10b}),
\begin{equation}\label{93}
F_{\textrm{Cas}}^{\textrm{Neu},\,l=0}(a)=-\frac{1}{2}W^{\prime\,l=0}_{I}\left(-\frac{1}{2},a\right)-\frac{1}{2}W^{\prime\,l=0}_{II}\left(-\frac{1}{2},a\right)-\frac{1}{2}\mathscr{F}_{\mathcal{N}}^{\prime\,l=0}\left(-\frac{1}{2},a,b\right)-\frac{d-1}{4\pi a^{2}}\;.
\end{equation}
Obviously, the total Casimir force on the piston for Neumann boundary conditions is the sum of (\ref{93}) and (\ref{68}) where in the latter the
lowest angular eigenvalue is omitted. In the next subsections we present explicit results for specific dimensions $d$. We would like to point
out that in the formulas that will follow it is understood that the functions $Z$, $W$ and $\mathscr{F}$ are evaluated for the specific dimension under consideration.

\subsection{Specific Dimensions for Dirichlet Boundary Conditions}

In the following special cases, we will set, for simplicity, $b=1$.
When the piston $\mathscr{N}$ is a sphere of dimension $d=2$, and, therefore, the dimension of $\mathscr{M}$ is $D=3$,
we obtain
\begin{equation}
F_{\textrm{Cas}}^{\textrm{Dir}}(a)=-\frac{1}{2}Z'_{I}\left(-\frac{1}{2},a\right)-\frac{1}{2}Z'_{II}\left(-\frac{1}{2},a\right)-\frac{1}{2}\mathscr{F}'_{D}\left(-\frac{1}{2},a\right)\;.
\end{equation}

For $d=3$, or $D=4$, we have the result
\begin{eqnarray}
F_{\textrm{Cas}}^{\textrm{Dir}}(a,\alpha)&=&-\frac{1}{2}Z'_{I}\left(-\frac{1}{2},a\right)-\frac{1}{2}Z'_{II}\left(-\frac{1}{2},a\right)-\frac{1}{2}\mathscr{F}'_{D}\left(-\frac{1}{2},a\right)
+\frac{1}{a^{2}}\Bigg(\frac{2803}{983040}+\frac{35}{65536}\gamma+\frac{35}{131072}\ln a^{2}\Bigg)\nonumber\\
&+&\frac{35}{131072a^{2}}\left(\frac{1}{\alpha}+\ln\mu^{2}\right)\;.
\end{eqnarray}

For $d=4$, or $D=5$, we obtain
\begin{eqnarray}
F_{\textrm{Cas}}^{\textrm{Dir}}(a)&=&-\frac{1}{2}Z'_{I}\left(-\frac{1}{2},a\right)-\frac{1}{2}Z'_{II}\left(-\frac{1}{2},a\right)-\frac{1}{2}\mathscr{F}'_{D}\left(-\frac{1}{2},a\right)
-\frac{35\pi^{2}}{1572864 a^{2}}\;.
\end{eqnarray}

And, finally, for $d=5$, or $D=6$, we get
\begin{eqnarray}
F_{\textrm{Cas}}^{\textrm{Dir}}(a,\alpha)&=&-\frac{1}{2}Z'_{I}\left(-\frac{1}{2},a\right)-\frac{1}{2}Z'_{II}\left(-\frac{1}{2},a\right)-\frac{1}{2}\mathscr{F}'_{D}\left(-\frac{1}{2},a\right)
+\frac{1}{a^{2}}\Bigg(\frac{20377}{113246208}-\frac{7285}{25165824}\gamma\nonumber\\
&-&\frac{1685}{50331648} \ln a^{2}+\frac{565}{25165824}\zeta_{R}(3)\Bigg)-\frac{1685}{50331648a^{2}}\left(\frac{1}{\alpha}+\ln\mu^{2}\right)\;.
\end{eqnarray}

\begin{figure}[h]
\centering
\mbox{\subfigure[\;$d=2$ ($D=3$)]{\includegraphics[width=3in]{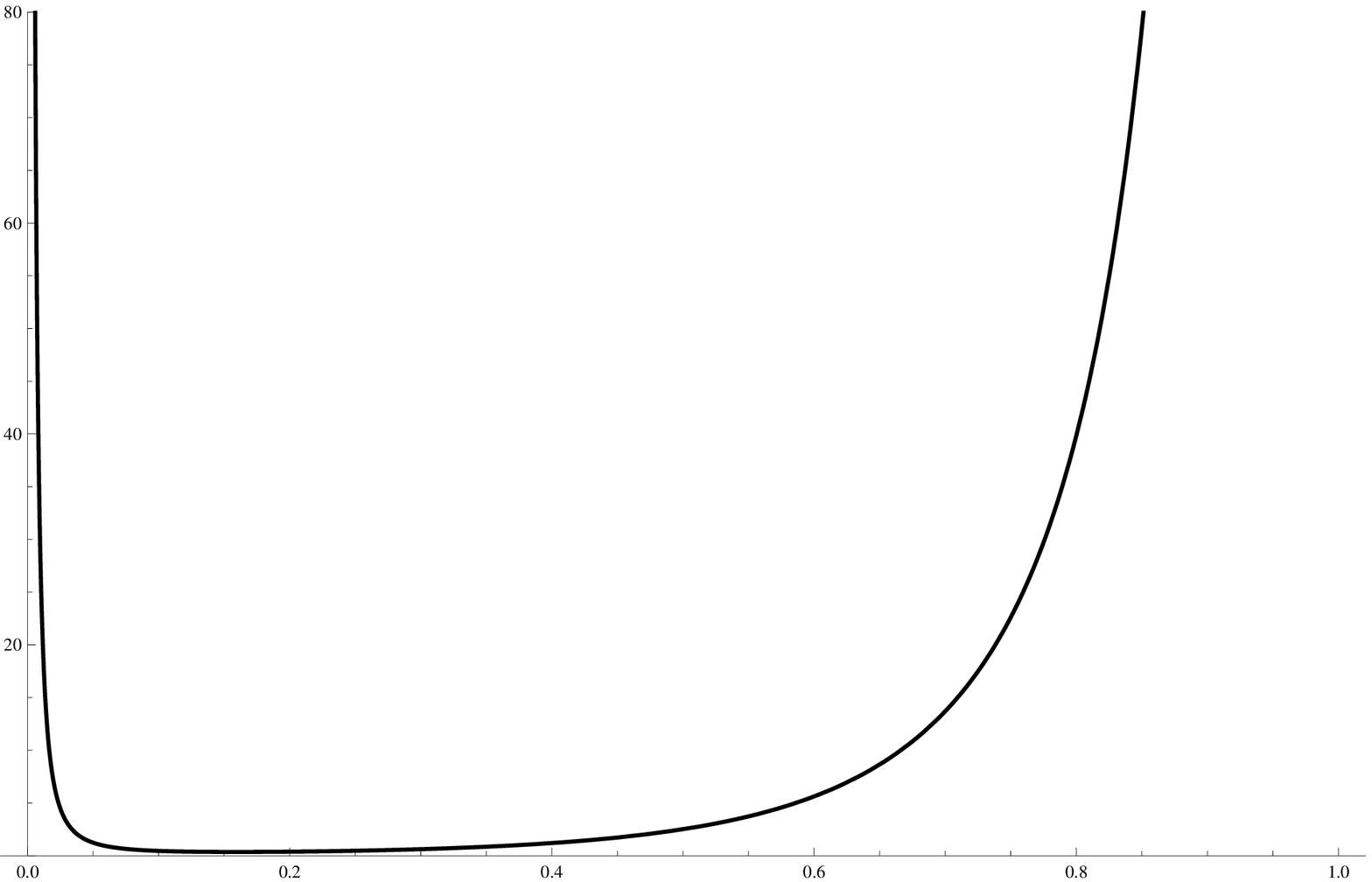}}\quad
\subfigure[\;$d=4$ ($D=5$)]{\includegraphics[width=3in]{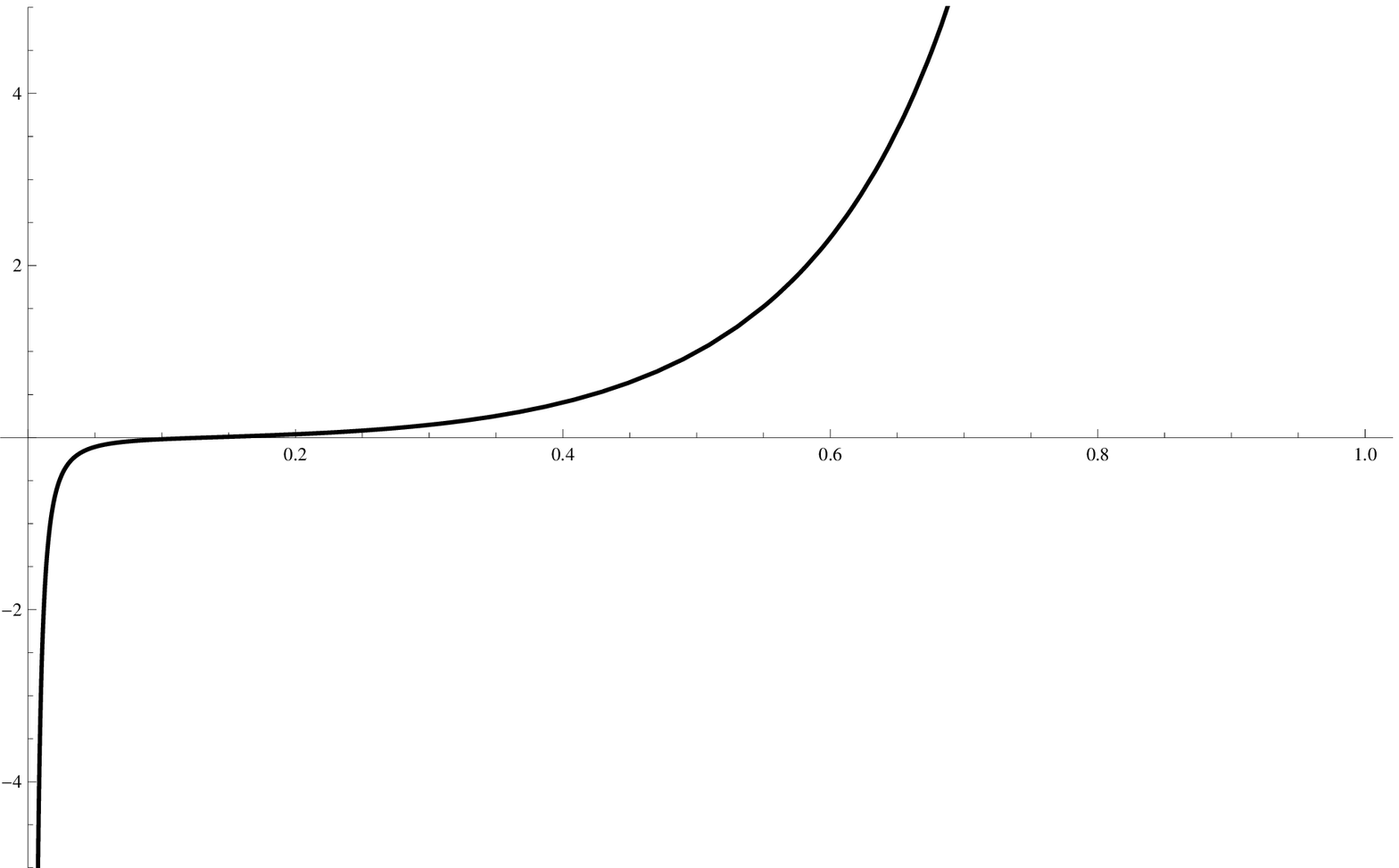} }}
\caption{Plots of the Casimir force, $F_{\textrm{Cas}}^{\textrm{Dir}}(a)$, on the piston $\mathscr{N}$ for Dirichlet boundary conditions as a function of the position $a$.} \label{fig1}
\end{figure}

As mentioned previously we can see, from the above results that the Casimir force on the piston is not a well defined quantity when the piston $\mathscr{N}$ is {\it odd} dimensional.
The Casimir force on the piston for $d=2$ and $d=4$, is shown in Figure \ref{fig1}. We can see that, for $d=2$, the piston is repelled from the conical singularity and attracted to the base manifold positioned at $b=1$. For $d=4$, instead, a point of unstable equilibrium is present. If the piston is to the left of this point, it is attracted to the conical
singularity. If it is to its right, then the piston is attracted to the base manifold.

\subsection{Specific Dimensions for Neumann Boundary Conditions}

For Neumann boundary conditions, we need to consider the results obtained in (\ref{68}), with the lowest angular eigenvalue omitted, and
the one obtained in (\ref{93}). It is useful to stress that the spectral $\zeta$-function $\bar{\zeta}_{\mathscr{N}}(s)$ with the lowest angular eigenvalue omitted is related to the ordinary $\zeta_{\mathscr{N}}(s)$ through the relation
\begin{equation}
\bar{\zeta}_{\mathscr{N}}(s)=\zeta_{\mathscr{N}}(s)-\left(\frac{d-1}{2}\right)^{-2s}\;.
\end{equation}
This expression is utilized in the results that will follow in order to obtain the Neumann Casimir force.

For $d=2$, thus $D=3$, we have the following expression for the force
\begin{eqnarray}
F_{\textrm{Cas}}^{\textrm{Neu}}(a)&=&-\frac{1}{2}W^{\prime\,l=0}_{I}\left(-\frac{1}{2},a\right)-\frac{1}{2}W^{\prime\,l=0}_{II}\left(-\frac{1}{2},a\right)-\frac{1}{2}\mathscr{F}_{\mathcal{N}}^{\prime\,l=0}\left(-\frac{1}{2},a\right)-\frac{1}{2}W'_{I}\left(-\frac{1}{2},a\right)\nonumber\\
&-&\frac{1}{2}W'_{II}\left(-\frac{1}{2},a\right)-\frac{1}{2}\mathscr{F}'_{\mathcal{N}}\left(-\frac{1}{2},a\right)-\frac{1}{a^{2}}\left(\frac{1}{4\pi}+\frac{27}{512}\right)\;.
\end{eqnarray}

For $d=3$, thus $D=4$, we have the result
\begin{eqnarray}
F_{\textrm{Cas}}^{\textrm{Neu}}(a,\alpha)&=&-\frac{1}{2}W^{\prime\,l=0}_{I}\left(-\frac{1}{2},a\right)-\frac{1}{2}W^{\prime\,l=0}_{II}\left(-\frac{1}{2},a\right)-\frac{1}{2}\mathscr{F}_{\mathcal{N}}^{\prime\,l=0}\left(-\frac{1}{2},a\right)-\frac{1}{2}W'_{I}\left(-\frac{1}{2},a\right)\nonumber\\
&-&\frac{1}{2}W'_{II}\left(-\frac{1}{2},a\right)-\frac{1}{2}\mathscr{F}'_{\mathcal{N}}\left(-\frac{1}{2},a\right)+\frac{1}{a^{2}}\left(-\frac{1}{2\pi}+\frac{12869}{32768}-\frac{5861}{65536}\gamma-\frac{5861}{131072}\ln a^{2}\right)\nonumber\\
&-&\frac{5861}{131072a^{2}}\left(\frac{1}{\alpha}+\ln\mu^{2}\right)\;.
\end{eqnarray}

For $d=4$, thus $D=5$, we obtain
\begin{eqnarray}
F_{\textrm{Cas}}^{\textrm{Neu}}(a)&=&-\frac{1}{2}W^{\prime\,l=0}_{I}\left(-\frac{1}{2},a\right)-\frac{1}{2}W^{\prime\,l=0}_{II}\left(-\frac{1}{2},a\right)-\frac{1}{2}\mathscr{F}_{\mathcal{N}}^{\prime\,l=0}\left(-\frac{1}{2},a\right)-\frac{1}{2}W'_{I}\left(-\frac{1}{2},a\right)\nonumber\\
&-&\frac{1}{2}W'_{II}\left(-\frac{1}{2},a\right)-\frac{1}{2}\mathscr{F}'_{\mathcal{N}}\left(-\frac{1}{2},a\right)+\frac{1}{a^{2}}\Bigg(-\frac{3}{4\pi}+\frac{57781}{221184}+\frac{27253}{1572864}\pi^{2}\Bigg)\;.
\end{eqnarray}

And, finally, for $d=5$, thus $D=6$, we get

\begin{eqnarray}
F_{\textrm{Cas}}^{\textrm{Neu}}(a,\alpha)&=&-\frac{1}{2}W^{\prime\,l=0}_{I}\left(-\frac{1}{2},a\right)-\frac{1}{2}W^{\prime\,l=0}_{II}\left(-\frac{1}{2},a\right)-\frac{1}{2}\mathscr{F}_{\mathcal{N}}^{\prime\,l=0}\left(-\frac{1}{2},a\right)-\frac{1}{2}W'_{I}\left(-\frac{1}{2},a\right)\nonumber\\
&-&\frac{1}{2}W'_{II}\left(-\frac{1}{2},a\right)-\frac{1}{2}\mathscr{F}'_{\mathcal{N}}\left(-\frac{1}{2},a\right)+\frac{1}{a^{2}}\Bigg(-\frac{1}{\pi}+\frac{53466379829}{126835752960}-\frac{1723783}{16777216}\gamma\nonumber\\
&-&\frac{1723783}{33554432}\ln a^{2}+\frac{10381781}{50331648}\zeta_{R}(3)\Bigg)-\frac{1723783}{33554432a^{2}}\left(\frac{1}{\alpha}+\ln\mu^{2}\right)\;.
\end{eqnarray}

\begin{figure}[h]
\centering
\mbox{\subfigure[\;$d=2$, and $D=3$]{\includegraphics[width=3in]{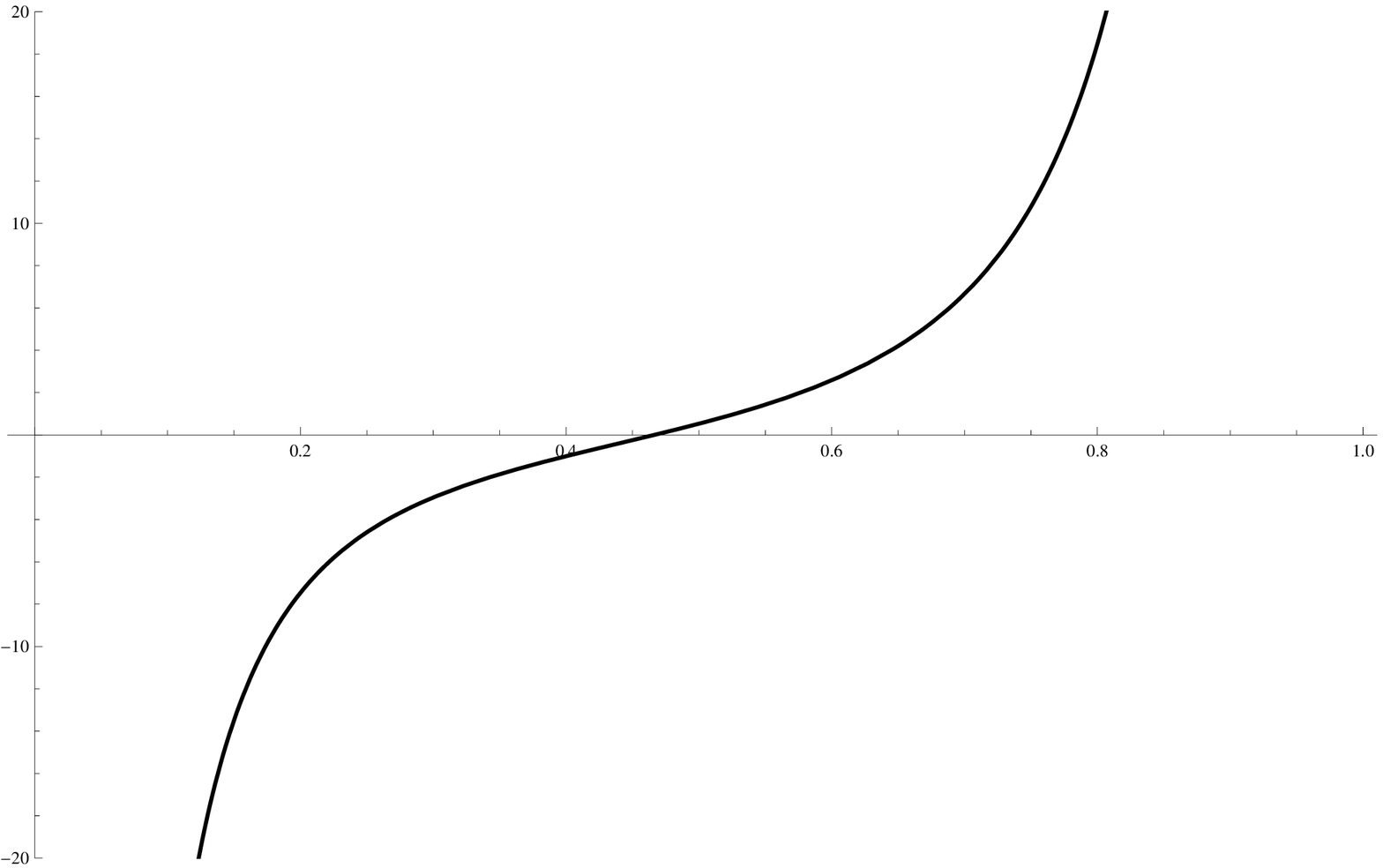}}\quad
\subfigure[\;$d=4$, and $D=5$]{\includegraphics[width=3in]{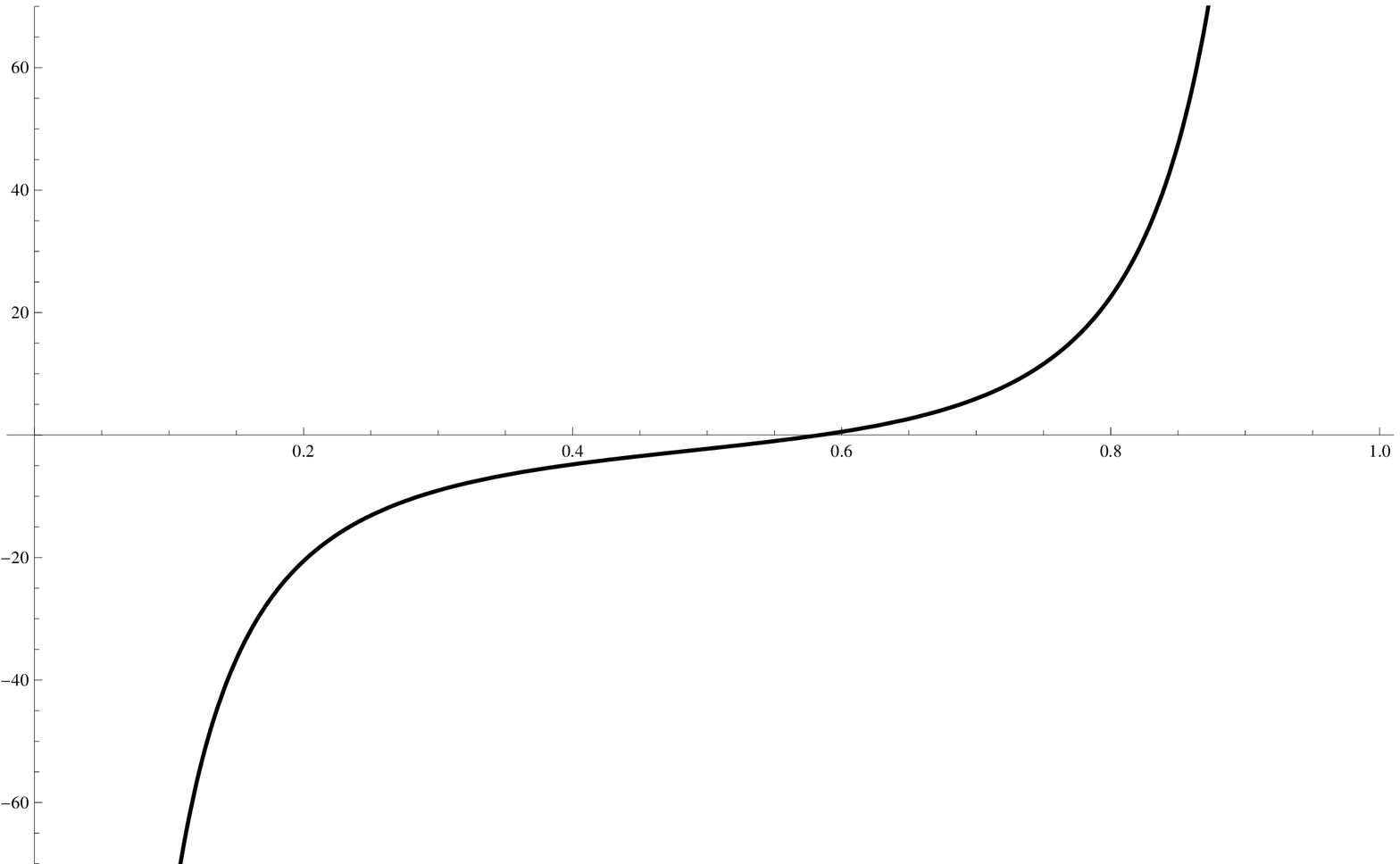} }}
\caption{Plots of the Casimir force, $F_{\textrm{Cas}}^{\textrm{Neu}}(a)$, on the piston $\mathscr{N}$ for Neuman boundary conditions as function of the position $a$.} \label{fig2}
\end{figure}

Once again, we would like to point out that also for Neumann boundary conditions the Casimir force on $\mathscr{N}$ is not well defined when $d$ is odd.
The Casimir force on the piston for $d=2$ and $d=4$, is shown in Figure \ref{fig2}. We can notice that in both cases, namely $d=2$ and $d=4$, there exists a point of unstable equilibrium. If the piston is to the left of this point it is attracted to the conical singularity, while, if it is on its right, it is attracted to the base manifold.

\section{Concluding Remarks}

In this paper we have analyzed the Casimir effect for massless scalar fields in the setting of the bounded generalized cone in the cases in which the field is endowed with
either Dirichlet or Neuman boundary conditions. The piston geometry has been obtained by dividing the generalized cone $\mathscr{M}$ into two regions separated by
its cross section positioned at $a$ with $a\in(0,1)$. The cross section itself represents the piston, and the structure so obtained has been denoted conical piston.
By utilizing $\zeta$-function regularization methods, we have obtained very general results for the Casimir energy and, hence, for the corresponding Casimir force.
The expressions obtained for the Casimir force are valid for any piston $\mathscr{N}$ and any dimension $D$ and are written in terms of the spectral $\zeta$-function
$\zeta_{\mathscr{N}}$. In order to obtain more specific results, we have considered in detail the case when the manifold $\mathscr{N}$ is a $d$-dimensional sphere.
Explicit numerical results for this case have also been obtained for $d=2$ and $d=4$.

We would like to stress that this work provides a study of a piston geometry that is more general than the ones considered in the literature where the two chambers have fundamentally the same type of geometry. The two separate regions of the conical piston, instead, have different geometries since one of them contains a singularity while the other does not. This is the feature that makes the conical piston particularly interesting.
These types of conical pistons are important, e.g., for the study of the interaction, through the Casimir force, of a piston
and a singular point (such as the tip of the cone). This results may also have some relevance in
in the framework of field theories with orbifold compactification.

The next step in this kind of investigations would be to analyze the Casimir force for massless scalar fields
on the conical piston satisfying mixed, or hybrid, boundary conditions. It would be interesting to further understand how the Casimir force
between the piston and the singular point of the manifold $\mathscr{M}$ is influenced by the boundary conditions imposed. Furthermore, it would be desirable to have a deeper understanding
of the influence of the geometry on ${\mathscr{N}}$ on the force.


\begin{acknowledgments}
The authors are grateful to Matthew Beauregard for his help in obtaining the plots. KK is supported by the National Science Foundation Grant
PHY-0757791.
\end{acknowledgments}

\appendix

\section{Polynomials $D_{n}(t)$ and $M_{n}(t,\beta )$ up to the order $n=6$}\label{app1}
In this appendix we list for completeness the polynomials $D_{n}(t)$ and $M_n (t,\beta )$ up to the sixth order.

By utilizing the cumulant expansion (\ref{23})
and the recurrence relation (\ref{18}) one obtains
\begin{eqnarray}
D_{1}(t)&=&\frac{1}{8}t-\frac{5}{24}t^{3}\;,\label{A1}\\
D_{2}(t)&=&\frac{1}{16}t^2-\frac{3 }{8}t^4+\frac{5 }{16}t^6\;,\\
D_{3}(t)&=&\frac{25 }{384}t^3-\frac{531 }{640}t^5+\frac{221 }{128}t^7-\frac{1105 }{1152}t^9\;,\\
D_{4}(t)&=&\frac{13}{128}t^4-\frac{71 }{32}t^6+\frac{531 }{64}t^8-\frac{339}{32}t^{10}+\frac{565 }{128}t^{12}\;,\\
D_{5}(t)&=&\frac{1073}{5120}t^5-\frac{50049 }{7168}t^7+\frac{186821 }{4608}t^9-\frac{44899 }{512}t^{11}+\frac{82825 }{1024}t^{13}-\frac{82825 }{3072}t^{15}\;,\\
D_{6}(t)&=&\frac{103}{192}t^6-\frac{405 }{16}t^8+\frac{1677 }{8}t^{10}-\frac{5389 }{8}t^{12}+\frac{65385 }{64}t^{14}-\frac{11805 }{16}t^{16}+\frac{19675 }{96}t^{18}\;.
\end{eqnarray}

The polynomials $M_{n}(t,\beta )$ follow by using  (\ref{51}) together with the recurrence relation (\ref{50}).
One can find
\begin{eqnarray}
M_{1}(t,\beta)&=&-\frac{3}{8} t+\frac{7}{24} t^3+t \beta\;,\label{A7}\\
M_{2}(t,\beta)&=&-\frac{3}{16}t^2+\frac{5}{8}t^4-\frac{7}{16}t^6+\frac{\beta}{2}t^2-\frac{\beta}{2}t^4-\frac{\beta^2}{2}t^2\;,\\
M_{3}(t,\beta)&=&-\frac{21}{128} t^3+\frac{869}{640} t^5-\frac{315}{128} t^7+\frac{1463}{1152} t^9+\frac{3\beta}{8} t^3 -\frac{5\beta}{4} t^5 +\frac{7\beta}{8} t^7 -\frac{\beta^2}{2}t^3\nonumber\\
&+&\frac{\beta^2}{2}t^5 +\frac{\beta^3}{3}t^3 \;,\\
M_{4}(t,\beta)&=&-\frac{27}{128} t^4+\frac{109}{32} t^6-\frac{733}{64} t^8+\frac{441}{32} t^{10}-\frac{707}{128} t^{12}
+\frac{3\beta}{8} t^4 -\frac{23\beta}{8} t^6 +\frac{41\beta}{8} t^8\nonumber\\
&-&\frac{21\beta}{8} t^{10} -\frac{\beta^2}{2}t^4
+\frac{3\beta^2}{2} t^6 -t^8 \beta^2+\frac{\beta^3}{2}t^4 -\frac{\beta^3}{2}t^6 -\frac{\beta^4}{4}t^4 \;,
\end{eqnarray}
\begin{eqnarray}
M_{5}(t,\beta)&=&-\frac{1899}{5120} t^5+\frac{72003}{7168} t^7-\frac{247735}{4608} t^9+\frac{56761}{512} t^{11}-\frac{101395}{1024} t^{13}
+\frac{495271}{15360} t^{15}+\frac{63\beta}{128} t^5\nonumber\\
&-&\frac{233\beta}{32} t^7 +\frac{1537\beta}{64} t^9 -\frac{917\beta}{32} t^{11} +\frac{1463\beta}{128} t^{13}
-\frac{9\beta^2}{16} t^5 +\frac{59\beta^2}{16} t^7 -\frac{99\beta^2}{16} t^9 +\frac{49\beta^2}{16} t^{11}\nonumber\\
&+&\frac{5\beta^3}{8} t^5 -\frac{7\beta^3}{4} t^7
+\frac{9\beta^3}{8} t^9 -\frac{\beta^4}{2}t^5 +\frac{\beta^4}{2}t^7 +\frac{\beta^5}{5}t^5\;,\\
M_{6}(t,\beta)&=&-\frac{27}{32} t^6+\frac{69}{2} t^8-\frac{17163}{64} t^{10}+\frac{4973}{6} t^{12}-\frac{9789}{8} t^{14}
+\frac{3465}{4} t^{16}-\frac{45493}{192} t^{18}+\frac{27\beta}{32} t^6 \nonumber\\
&-&\frac{681\beta}{32} t^8 +\frac{1793\beta}{16} t^{10}
-\frac{3671\beta}{16} t^{12} +\frac{6531\beta}{32} t^{14} -\frac{2121\beta}{32} t^{16} -\frac{3\beta^2}{4} t^6 +\frac{75\beta^2}{8} t^8 \nonumber\\
&-&\frac{233\beta^2}{8} t^{10} +\frac{269\beta^2}{8} t^{12} -\frac{105\beta^2}{8} t^{14} +\frac{19\beta^3}{24} t^6 -\frac{37\beta^3}{8} t^8
+\frac{59\beta^3}{8} t^{10} -\frac{85\beta^3}{24} t^{12} \nonumber\\
&-&\frac{3\beta^4}{4} t^6 +2 t^8 \beta^4-\frac{5\beta^4}{4} t^{10} +\frac{\beta^5}{2}t^6
-\frac{\beta^5}{2}t^8 -\frac{\beta^6}{6}t^6 \;.
\end{eqnarray}

\end{document}